\documentclass
[superscriptaddress,secnumarabic,amssymb,amsmath,nobibnotes,aps,prd,showkeys,showpacs,nofootinbib,onecolumn,12pt]{revtex4}%
\usepackage{graphicx}
\usepackage{epsf}
\usepackage{bm}
\usepackage{amsmath}
\usepackage{amsfonts}
\usepackage{amssymb}
\usepackage{color}%
\providecommand{\U}[1]{\protect\rule{.1in}{.1in}}

\newcommand{\be}{\begin{equation}}
\newcommand{\ee}{\end{equation}}

\newcommand{\mincir}{\raise
-3.truept\hbox{\rlap{\hbox{$\sim$}}\raise4.truept\hbox{$<$}\ }}
\newcommand{\magcir}{\raise
-3.truept\hbox{\rlap{\hbox{$\sim$}}\raise4.truept\hbox{$>$}\ }}

\usepackage{tikz,xcolor,hyperref}
\hypersetup{
    colorlinks=true,
    linkcolor=red,
    citecolor=blue,
    filecolor=magenta,      
    urlcolor=cyan,
    pdftitle={Averaging generalized scalar field cosmologies IV},
    pdfpagemode=FullScreen}
    
\begin{document}
\title{Dynamical system analysis in multiscalar-torsion cosmology}
\author{Genly Leon }
\email{genly.leon@ucn.cl}
\affiliation{Departamento de Matem\'{a}ticas, Universidad Cat\'{o}lica del Norte, Avda.
Angamos 0610, Casilla 1280 Antofagasta, Chile}
\affiliation{Institute of Systems Science, Durban University of Technology, PO Box 1334,
Durban 4000, South Africa}
\author{Andronikos Paliathanasis }
\email{anpaliat@phys.uoa.gr}
\affiliation{Institute of Systems Science, Durban University of Technology, PO Box 1334,
Durban 4000, South Africa}
\affiliation{Departamento de Matem\'{a}ticas, Universidad Cat\'{o}lica del Norte, Avda.
Angamos 0610, Casilla 1280 Antofagasta, Chile}
\author{Alfredo D. Millano}
\email{alfredo.millano@alumnos.ucn.cl}
\affiliation{Departamento de Matem\'{a}ticas, Universidad Cat\'{o}lica del Norte, Avda.
Angamos 0610, Casilla 1280 Antofagasta, Chile}
\begin{abstract}
We explore the phase-space of a multiscalar-torsion gravitational theory
within a cosmological framework characterized by a spatially flat
Friedmann--Lema\^{\i}tre--Robertson--Walker model. Our investigation focuses
on teleparallelism and involves a gravitational model featuring two scalar
fields, where one scalar field is coupled to the torsion scalar. We consider
coupling in the two scalar fields' kinetic and potential components. We employ
exponential functions for the scalar field potentials and analyze the field
equations' equilibrium points to reconstruct the cosmological evolution.
Remarkably, we discover many equilibrium points in this multiscalar field
model, capable of describing various eras of cosmological evolution. Hence,
this model can be used to describe the early and late time acceleration phases
of the universe and as a unification model for the elements of the dark sector
of the universe.

\end{abstract}
\keywords{Teleparallel; Scalar field; Scalar-torsion; Dynamical analysis}
\pacs{98.80.-k, 95.35.+d, 95.36.+x}
\date{\today}
\maketitle

\section{Introduction}

\label{sec1}

Recent cosmological observations suggest that our universe is currently
experiencing acceleration \cite{rr1, Teg, Kowal, Komatsu, suzuki11}, as well
as having undergone a period of acceleration during its early stage known as
cosmic inflation \cite{planckinflation}. The early acceleration phase is
attributed to the inflation \cite{guth}, while dark energy is believed to be
responsible for the late-time acceleration. Both dark energy and inflaton are
fluid sources with negative pressure components that provide acceleration. A
fascinating possibility is that these two eras of acceleration, inflation and
late-time acceleration, may be interconnected. It is conceivable that dark
energy and inflation arise from the same underlying physical mechanism,
manifesting at different stages in the universe's evolution. However, a
comprehensive theory needs to be revised to explain the connection between
inflation and late-time acceleration, particularly considering the vast
disparity in energy scales between these two phenomena.

The concept of cosmic acceleration is often explained through scalar fields.
In the context of inflation, a scalar field with a nonzero mass term is
assumed to drive the dynamics, resulting in an acceleration period during the
slow-roll limit. Various potential functions for scalar fields have been
proposed in the literature for inflation, as seen in references
\cite{ref1,ref2,newinf,ref4,ref6,ref7,ref8}. A review article, reference
\cite{ref9}, provides a detailed list of these proposed inflationary models.
The quintessence theory is a straightforward model for dynamical dark energy
\cite{Ratra, Barrow}. This theory involves introducing an adjusting or tracker
scalar field \cite{Caldwell98}, which, while rolling down the potential
function, can provide acceleration \cite{qq1,qq2,qq3,qq4,qq5,qq6,q15,q16}. The
quintessence model can also describe the degrees of freedom of other dark
energy models, such as the Chaplygin gas fluids \cite{q18,q19,q20}, to provide
a unified description for the dark energy components of the universe
\cite{q17,q21}.

There are several proposed generalizations of scalar field models. One such
generalization is the phantom scalar field \cite{pha1,pha2,pha3}. In this
model, the presence of a scalar field allows for the possibility of negative
energy density. As a consequence, the equation of the state parameter can
cross the phantom divide line. That leads to an accelerated universe expansion
that exceeds the exponential growth predicted by the de Sitter solution,
ultimately resulting in a cosmological singularity known as the Big Rip
\cite{q22}. \ Another proposal to explain these phenomena is the chameleon
theory, which suggests that the scalar field interacts with other matter
components in the universe. This interaction gives rise to a chameleon
mechanism, whereby the scalar field's properties change depending on the
surrounding matter environment.\cite{ch1,ch2}. In this theory, the mass of the
scalar field depends on the mass of the other matter components of the
universe \cite{salim96,ame1}.

Extensive research has been conducted on nonminimally coupled scalar fields,
which involve the interaction between the scalar field and gravity within the
gravitational action integral. These are known as scalar-tensor theories of
gravity \cite{faraonibook} and include the Brans-Dicke theory as a particular
case \cite{Brans}. These theories are consistent with Mach's principle, which
requires the existence of the scalar field. They are related to minimally
coupled scalar field models by a geometric map that connects the Einstein and
the Jordan frames through a conformal transformation \cite{bb00a}. However,
the physical properties of a gravitational theory rarely survive this
transformation \cite{bb02,bb03}. Other nonminimally coupled scalar field
gravitational theories have also been proposed \cite{sf1,sf2,sf3,sf4,sf5}.
Multiscalar field models have also been studied extensively
\cite{qu1,qu2,qu3,qu4} as they provide new degrees of freedom to explain
various eras of cosmological history and connect the early-time acceleration
phase with the late-time acceleration \cite{qu5}.

Modified gravity theories offer alternative explanations for the universe's
acceleration \cite{mod1, mod2}. These theories modify the Einstein-Hilbert
Action by introducing geometric invariants in the field equations that provide
acceleration through geometrodynamical terms. In General Relativity, the Ricci
scalar $R$ is the fundamental scalar constructed using the Levi-Civita
connection. However, using a general connection, we can construct the
non-metricity scalar $Q$ and the torsion scalar $T$, which, along with $R$,
are known as the trinity of gravity \cite{tr1}. When the antisymmetric
Weitzenb\"{o}ck connection is the fundamental connection of physical space
\cite{Weitzenb23}, the torsion scalar $T$ can be used as gravitational
Lagrangian \cite{cc,Hayashi79}, leading to the teleparallel equivalent of
General Relativity. Introducing a nonminimally coupled scalar field results in
a scalar-torsion theory \cite{sss1, ss2, pal1}, which differs from the
scalar-tensor theory of gravity \cite{pal2}. The scalar-torsion theory has
been applied to describe the late-time acceleration of the universe in
\cite{te1, te3, te5,dd0}. From the phase-space analysis of the field
equations in scalar-torsion theory \cite{dd0a} it was found that the
asymptotic solutions described by the quintessence scalar field or the phantom
scalar field of General Relativity are provided. Additionally, the de Sitter
point always exists. In \cite{dd1}, inflation in scalar-tensor theory was
investigated. The analysis of the dynamics found that some asymptotic saddle points can be attributed to inflation.\ For a discussion we refer the reader to \cite{dd2,dd3,dd4,dd5,dd6}
and references therein.

A multiscalar field cosmology was recently explored in
teleparallelism in \cite{dial}. In this model, one of the fields is coupled to
the torsion scalar $T$ through a diatonic coupling function, while
the other field minimally couples to gravity. Still, there exists nonzero
interaction between the two fields. The field equations for the multiscalar
field models are of second order, and the Noether symmetry analysis is used to
determine exact and analytic solutions and specify the scalar field
potentials.

This research examines the multiscalar-torsion theory in the publication by
\cite{dial}. Our goal is to conduct a thorough phase-space analysis that will allow
us to reconstruct the universe's history and evaluate the feasibility of the
gravitational model proposed in \cite{da1}. This analytical approach has been
widely utilized in various gravitational theories, as evident in the
references \cite{da2,da3,da4,da5,da6,da7}. Using phase-space analysis, we can
effectively identify crucial epochs in cosmological history and draw
insightful conclusions regarding the initial value problem. The paper is
structured as follows.

Section \ref{sec2}  introduces teleparallelism's
fundamental concepts and definitions. We also discuss our gravitational
model, which involves two coupled scalar fields within scalar-torsion theory.
Our primary focus will be on studying a spatially homogeneous cosmology, where
we will derive the modified Friedmann equations for the
Friedmann-Lema\^{\i}tre-Robertson-Walker metric tensor. To investigate the
overall evolution of the physical parameters described by the modified
Friedmann equations, we will introduce dimensionless variables and express the
cosmological field equations as first-order algebraic-differential equations
in Section \ref{sec3}. The novel contributions of this study are covered in
Sections \ref{sec4} and \ref{sec5}, where we present the equilibrium points of
the field equations in both finite and infinite regimes. Each equilibrium point
represents an asymptotic solution for the cosmological model. We will delve
into the physical properties of these solutions and analyze their stability,
as this information is crucial for understanding the cosmological evolution of
the multiscalar-torsion model and addressing the theory's initial value
problem. Finally, in Section \ref{conc}, we will summarise our findings.

\section{Multiscalar-torsion gravity}

\label{sec2}

We work in the context of the teleparallel theory of gravity coupled
to a scalar field. In particular, we consider the multiscalar field gravitational model with Action Integral \cite{te1}%
\begin{equation}
S=\frac{1}{16\pi G}\int\mathrm{d}^{4}xe\left[  F\left(  \phi\right)  \left(
T+\frac{\omega}{2}\phi_{;\mu}\phi^{\mu}+U\left(  \phi\right)  +L\left(
x^{\kappa},\phi,\psi,\psi_{;\mu}\right)  \right)  \right], \label{d.08}%
\end{equation}
which belongs to the family of scalar-torsion theories.

Where in the Action Integral \eqref{d.08}, $\phi$ is the
scalar field nonminimally coupled to the torsion scalar $T$,
$F\left(  \phi\right)  $ is the coupling function between the scalar
field $\phi$  and  $T$, $U\left(  \phi\right)  $ is
the scalar field potential, and$L\left(  x^{\kappa},\phi,\psi,\psi_{;\mu
}\right)  $ describe the\ dynamics of a second-scalar field. $\psi_{;\mu
}$ denotes the covariant derivative. Parameter $\omega$ is arbitrary, and its value reflects the coupling with
gravity.

The torsion scalar $T$ is defined by the antisymmetric
Weitzenb\"{o}ck connection $\hat{\Gamma}^{\lambda}{}_{\mu\nu}$, that
is,
\begin{equation}
T={S_{\beta}}^{\mu\nu}{T^{\beta}}_{\mu\nu},
\end{equation}
in which $T_{\mu\nu}^{\beta}$ is the torsion tensor defined
as 
\begin{equation}
T_{\mu\nu}^{\beta}=\hat{\Gamma}_{\nu\mu}^{\beta}-\hat{\Gamma}_{\mu\nu}^{\beta
},
\end{equation}
and%
\begin{equation}
{S_{\beta}}^{\mu\nu}=\frac{1}{2}({K^{\mu\nu}}_{\beta}-\delta_{\beta}^{\mu}%
{T}_{\theta}^{~~\theta\nu}+\delta_{\beta}^{\nu}{T}_{\theta}^{~~\theta\mu}),
\end{equation}
$\ $ where now %
\begin{equation}
{K^{\mu\nu}}_{\beta}=-\frac{1}{2}({T^{\mu\nu}}_{\beta}-{T^{\nu\mu}}_{\beta
}-{T_{\beta}}^{\mu\nu}),
\end{equation}
is the contorsion tensor. The Weitzenb\"{o}ck connection $\hat
{\Gamma}^{\lambda}{}_{\mu\nu}$ is related to the vierbein fields
$e_{i}=h_{i}^{\mu}\partial_{i}$, as $\hat{\Gamma}^{\lambda}{}%
_{\mu\nu}=h_{a}^{\lambda}\partial_{\mu}h_{\nu}^{a}+\omega_{~\nu\kappa}%
^{a}h_{\mu}^{\kappa}h_{a}^{\lambda}$ with the metric tensor%
$g_{\mu\nu}~$ to be defined as $g_{\mu\nu}=\eta_{ij}h_{\mu}^{i}h_{\nu
}^{j}$,~and $\omega_{~\nu\kappa}^{\lambda}$ is the spin
connection \cite{spin1}.

For the second scalar field we assume $L\left(  x^{\kappa},\phi,\psi
,\psi_{;\mu}\right)  =\frac{\beta}{2}\psi_{;\mu}\psi^{;\mu}+\frac{\hat
{V}\left(  \phi,\psi\right)  }{F\left(  \phi\right)  }$, where now the Action
Integral \eqref{d.08} reads \cite{dial}
\begin{equation}
S=\frac{1}{16\pi G}\int\mathrm{d}^{4}xe\Big[F\left(  \phi\right)  \left(
T+\frac{\omega}{2}\phi_{;\mu}\phi^{\mu}+U\left(  \phi\right)  +\frac{\beta}%
{2}\psi_{;\mu}\psi^{;\mu}\right)  +\hat{V}\left(  \phi,\psi\right)  \Big],
\label{dd.01}%
\end{equation}
where $\beta^{2}=1$. We have introduced a new potential function $\hat
{V}\left(  \phi,\psi\right)  $ in order to be able to introduce interaction
between the two scalar fields $\phi$ and $\psi$. Parameters $\omega$ and
$\beta$ are nonzero parameters, and their values define the nature of the
scalar fields.

Last but not least, we want to discuss that in order to end the latter
Action (\ref{dd.01}) with starting point (\ref{d.08}), we considered a specific
representation for the scalar field; for more details, we refer the reader to
the discussion in \cite{scalar2}.

\subsection{FLRW cosmology}

In our analysis, we assume that the universe is described by the
four-dimensional isotropic, homogeneous, and spatially flat FLRW
(Friedmann-Lema\^{\i}tre-Robertson-Walker) geometry. This assumption allows us
to employ a specific line element to characterize the properties of the
universe. Specifically, we select the metric tensor $g_{\mu\nu}$ with line
element
\begin{equation}
\mathrm{d}s^{2}=-\mathrm{d}t^{2}+a^{2}(t)(\mathrm{d}x^{2}+\mathrm{d}%
y^{2}+\mathrm{d}z^{2}), \label{d.06}%
\end{equation}
where $a\left(  t\right)  $ is the scale factor.

Moreover, we consider the diagonal vierbein fields \cite{te1} %
\begin{equation}
h_{~\mu}^{i}(t)=\mathrm{diag}(1,a(t),a(t),a(t)),
\end{equation}
and the zero spin connection in Cartesian coordinates \cite{spin1}.

From the latter follows $T=6H^{2}$, with $H=\frac{\dot{a}}{a}$, $\dot
{a}=\frac{\mathrm{d}a}{\mathrm{d}t}$, to be the Hubble function. For the
latter vierbein fields, the limit of General Relativity is recovered when the
dynamical components of the scalar fields are eliminated.

Thus, from \eqref{dd.01} we calculate the point-like Lagrangian \cite{dial}
\begin{equation}
\mathcal{L}\left(  a,\dot{a},\phi,\dot{\phi},\psi,\dot{\psi}\right)  =F\left(
\phi\right)  \left(  6a\dot{a}^{2}+a^{3}\left(  \frac{\omega}{2}\dot{\phi}%
^{2}+\frac{\beta}{2}\dot{\psi}^{2}\right)  -a^{3}V\left(  \phi,\psi\right)
\right),  \label{dd.03}%
\end{equation}
while the gravitational field equations are
\begin{equation}
6H^{2}+\frac{\omega}{2}\dot{\phi}^{2}+\frac{\beta}{2}\dot{\psi}^{2}-V\left(
\phi,\psi\right)  =0, \label{dd.04}%
\end{equation}
\begin{equation}
\dot{H}+\frac{3}{2}H^{2}-\left(  \frac{1}{4}\left(  \frac{\omega}{2}\dot{\phi
}^{2}+\frac{\beta}{2}\dot{\psi}^{2}+V\left(  \phi,\psi\right)  \right)
-H\dot{\phi}\left(  \ln\left(  F\left(  \phi\right)  \right)  \right)
_{,\phi}\right)  =0, \label{dd.05}%
\end{equation}
\begin{equation}
\ddot{\phi}+3H\dot{\phi}+\frac{1}{\omega}\left(  \ln\left(  F\left(
\phi\right)  \right)  \right)  _{,\phi}\left(  \frac{\omega}{2}\dot{\phi}%
^{2}-\frac{\beta}{2}\dot{\psi}^{2}-6H^{2}-V\left(  \phi,\psi\right)  \right)
-\frac{1}{\omega}V_{,\phi}=0, \label{dd.06}%
\end{equation}
\begin{equation}
\ddot{\psi}+3H\dot{\psi}+\left(  \ln\left(  F\left(  \phi\right)  \right)
\right)  _{,\phi}\dot{\phi}\dot{\psi}-\frac{1}{\beta}V_{,\psi}=0.
\label{dd.07}%
\end{equation}
We have defined the function $V\left(  \phi,\psi\right)  $ such that $\hat
{V}\left(  \phi,\psi\right)  =-V\left(  \phi,\psi\right)  F\left(
\phi\right)  -F\left(  \phi\right)  U\left(  \phi\right)  .$

In the following, we consider the coupling function $F(\phi)=F_{0}e^{2\phi}$.
That is not an arbitrary selection. For such a function, in the absence of the
second-scalar field, the gravitational model admits an $O\left(  d,d\right)  $
duality symmetry and it has many similarities with the Brans-Dicke model of
scalar-tensor theory \cite{pal1,pal2}.

Therefore, for $F(\phi)=F_{0}e^{2\phi}$ the gravitational field equations
\eqref{dd.04}-\eqref{dd.07} read \cite{dial}
\begin{equation}
6H^{2}+\frac{\omega}{2}\dot{\phi}^{2}+\frac{\beta}{2}\dot{\psi}^{2}-V\left(
\phi,\psi\right)  =0, \label{dd.10}%
\end{equation}
\begin{equation}
\dot{H}+\frac{3}{2}H^{2}-\left(  \frac{1}{4}\left(  \frac{\omega}{2}\dot{\phi
}^{2}+\frac{\beta}{2}\dot{\psi}^{2}+V\left(  \phi,\psi\right)  \right)
-2H\dot{\phi}\right)  =0, \label{dd.11}%
\end{equation}
\begin{equation}
\ddot{\phi}+3H\dot{\phi}+\frac{2}{\omega}\left(  \frac{\omega}{2}\dot{\phi
}^{2}-\frac{\beta}{2}\dot{\psi}^{2}-6H^{2}-V\left(  \phi,\psi\right)  \right)
-\frac{1}{\omega}V_{,\phi}=0, \label{dd.12}%
\end{equation}%
\begin{equation}
\ddot{\psi}+3H\dot{\psi}+2\dot{\phi}\dot{\psi}-\frac{1}{\beta}V_{,\psi}=0.
\label{dd.13}%
\end{equation}

An equivalent way to write the field equations \eqref{dd.10}- \eqref{dd.11}
is
\begin{align}
3H^{2}  &  =\rho_{eff}, ~\\
2\dot{H}+3H^{2}  &  =-p_{eff},
\end{align}
where $\rho_{eff}$ and $p_{eff}$ are the components of the effective fluid
defined by the two scalar fields as follows
\begin{equation}
\rho_{eff}=\frac{1}{2}\left(  -\frac{\omega}{2}\dot{\phi}^{2}-\frac{\beta}%
{2}\dot{\psi}^{2}+V\left(  \phi,\psi\right)  \right), 
\end{equation}%
\begin{equation}
p_{eff}=\left(  \frac{1}{2}\left(  -\frac{\omega}{2}\dot{\phi}^{2}-\frac
{\beta}{2}\dot{\psi}^{2}-V\left(  \phi,\psi\right)  \right)  +4H\dot{\phi
}\right).
\end{equation}

The equation of state parameter for the effective fluid is defined as
$w_{eff}=\frac{p_{eff}}{\rho_{eff}}$, that is%
\begin{equation}
w_{eff}=\frac{\left(  -\frac{\omega}{2}\dot{\phi}^{2}-\frac{\beta}{2}\dot
{\psi}^{2}-V\left(  \phi,\psi\right)  \right)  +8H\dot{\phi}}{\left(
-\frac{\omega}{2}\dot{\phi}^{2}-\frac{\beta}{2}\dot{\psi}^{2}+V\left(
\phi,\psi\right)  \right)  }.
\end{equation}

In the limit $\dot{\phi}\rightarrow0$, we find
\begin{equation}
w_{eff}|_{\dot{\phi}\rightarrow0}\simeq\frac{-\frac{\beta}{2}\dot{\psi}%
^{2}-V\left(  \phi_{0},\psi\right)  }{-\frac{\beta}{2}\dot{\psi}^{2}+V\left(
\phi_{0},\psi\right)  },
\end{equation}
which means that a quintessence scalar describes the effective fluid field
$\left(  \beta<0\right)  $ or a phantom scalar field $\left(  \beta>0\right)
$.

On the other hand, in the limit $\dot{\phi}^{2}\rightarrow0$ and $\dot{\psi
}^{2}\rightarrow0$, we derive
\begin{equation}
w_{eff}|_{\left(  \dot{\phi}^{2},\dot{\psi}^{2}\right)  \rightarrow0}%
\simeq-1+\frac{8H\dot{\phi}}{V\left(  \phi,\psi\right)  },
\end{equation}
thus, the effective fluid can deviate from the cosmological constant and, for
$\frac{8H\dot{\phi}}{V\left(  \phi,\psi\right)  }<0$, $w_{eff}$ can cross the
phantom divide line.

Recently, in \cite{dial}, the integrability properties of the latter dynamical
system were studied using the Noether symmetry approach. Specifically, the
potential function $V\left(  \phi,\psi\right)  $ has been constrained by the
requirement for the field equations to admit Noether symmetries. As a result,
conservation laws. New exact and analytic solutions were determined. However,
in this work, we perform a detailed analysis of the phase space for the
dynamical system \eqref{dd.10}-\eqref{dd.13}. Hence, from such analysis, we
can reconstruct the cosmological history provided by the multiscalar-torsion
model and study the stability properties of the exact solutions determined in
\cite{dial}.

\section{Field equations in dimensionless variables}

\label{sec3}

We proceed by assuming the scalar field potential to be $V\left(  \phi
,\psi\right)  $ to be separable, that is $V\left(  \phi,\psi\right)
=V_{1}\left(  \phi\right)  +V_{2}\left(  \psi\right)  $. Consequently,
$\hat{V}\left(  \phi,\psi\right)  =-V_{2}\left(  \psi\right)  F\left(
\phi\right)  -F\left(  \phi\right)  \left(  U\left(  \phi\right)
+V_{1}\left(  \phi\right)  \right)  $, which means that there exist
interaction between the two scalar fields as long as $V_{2}\left(
\psi\right)  $ is a nonconstant function.

In the context of the $H$-normalization \cite{dn1} we define the dimensionless
variables
\begin{equation}
x=\frac{\dot{\phi}}{\sqrt{12}H}, ~y=\sqrt{\frac{V_{1}\left(  \phi\right)
}{6H^{2}}}, ~z=\frac{\dot{\psi}}{\sqrt{12}H}, ~v=\sqrt{\frac{V_{2}\left(
\psi\right)  }{6H^{2}}},
\end{equation}%
\begin{equation}
\lambda=\left(  \ln V_{1}\left(  \phi\right)  \right)  _{,\phi}, ~\mu=\left(
\ln V_{2}\left(  \psi\right)  \right)  _{,\psi}, ~\tau=\ln a\text{~}.
\end{equation}

In the new variables, the field equations \eqref{dd.10}-\eqref{dd.13} are
written in the equivalent form of an algebraic-differential system of
first-order differential equations.

The new dimensionless variables satisfy the following system
\begin{align}
2\omega\frac{dx}{d\tau} &=v^2 \left(4 \sqrt{3}-3 \omega  x\right)+\left(4 \sqrt{3}-3 \omega  x\right) \left(\omega 
   x^2+y^2+\beta  z^2+1\right)+2 \sqrt{3} \lambda  y^2,
\label{dd.20}%
\\
\frac{dy}{d\tau}&=-\frac{1}{2} y \left(3 v^2-2 \sqrt{3} (\lambda +4) x+3 \omega  x^2+3 \left(y^2+\beta 
   z^2-1\right)\right), \label{dd.22}%
\\
2\beta\frac{dz}{d\tau}&=v^2 \left(2 \sqrt{3} \mu -3 \beta  z\right)-3 \beta  z \left(\omega  x^2+y^2+\beta 
   z^2+1\right), \label{dd.21}%
\\
\frac{dv}{d\tau}&=-\frac{1}{2} v \left(3 v^2+3 \omega  x^2-8 \sqrt{3} x+3 y^2+3 \beta  z^2-2 \sqrt{3} \mu 
   z-3\right),
\label{dd.23}%
\\
\frac{d\lambda}{d\tau}&=2\sqrt{3}x\lambda^{2}\left(  \Gamma_{1}\left(
\lambda\right)  -1\right), \label{dd.24}%
\\
\frac{d\mu}{d\tau}&=2\sqrt{3}z\mu^{2}\left(\Gamma_{2}\left(  \mu\right)
-1\right), \label{dd.25}%
\end{align}
with algebraic constraint
\begin{equation}
1+\omega x^{2}+\beta z^{2}-y^{2}-v^{2}=0. \label{dd.26}%
\end{equation}
Functions $\Gamma_{1}\left(  \lambda\right)  $ and $\Gamma_{2}\left(
\mu\right)  $ are defined as
\begin{equation}
\Gamma_{1}\left(\lambda\right)  =\frac{V_{1,\phi\phi}V_{1}}{\left(
V_{1,\phi}\right)  ^{2}}\text{ and }\Gamma_{2}\left(  \mu\right)
=\frac{V_{2,\psi\psi}V_{2}}{\left(  V_{2,\psi}\right)  ^{2}}. \label{dd.27}%
\end{equation}

With the constraint equation \eqref{dd.26}, the dimension of the dynamical
system is reduced to five. However, for exponential potential $V_{1}\left(
\phi\right)  =V_{10}e^{\lambda\phi}$ and $V_{2}\left(  \psi\right)
=V_{20}e^{\mu\psi}$, we obtain constant values $\lambda$ and $\mu$.

In the
following from \eqref{dd.25} we derive $v=\sqrt{1+\omega x^{2}+\beta
z^{2}-y^{2}}$.

Thus, the dimension of the dynamical system is reduced to three:  
\begin{align}
\label{eq-a}\frac{dx}{d\tau}& =\frac{\sqrt{3} \left(\lambda  y^2+4 \beta  z^2+4\right)}{\omega }-x \left(3 \omega  x^2-4\sqrt{3} x+3 \beta  z^2+3\right), \\
\label{eq-b}\frac{dy}{d\tau}& = y \left(\sqrt{3} (\lambda +4) x-3 \omega  x^2-3 \beta z^2\right),\\
\label{eq-c}\frac{dz}{d\tau}& =\frac{\sqrt{3} \mu  \left(\omega  x^2-y^2+\beta  z^2+1\right)}{\beta }-3 z\left(\omega  x^2+\beta  z^2+1\right).
\end{align}

We proceed with the investigation of the equilibrium points for the
three-dimensional dynamical system \eqref{eq-a}, \eqref{eq-b} and \eqref{eq-c}. Every equilibrium point describes an
asymptotic exact cosmological solution with an effective equation of state
parameter%
\begin{equation}
w_{eff}\left(  x,y,z;\omega,\beta\right)  =-1-2\left(  \omega x^{2}+\beta
z^{2}\right)  +\frac{8}{\sqrt{3}}x.
\end{equation}
The deceleration parameter is given by 
\begin{equation}
q\left(  x, z;\omega,\beta\right)= -1   -3 \omega  x^2+4 \sqrt{3} x-3 \beta  z^2. 
\end{equation}

For $w_{eff}\neq-1$, the asymptotic solutions have a power-law scale factor $a\left(  t\right)  =a_{0}t^{\frac{2}{3\left(  1+w_{eff}\right)  }}$, while
for $w_{eff}=-1$, the universe is described by the de Sitter solution with $a\left(  t\right)  =a_{0}e^{H_{0}t}$. For the equilibrium points, we study their stability properties to construct the cosmological history. An equilibrium point characterized as an attractor indicates that the asymptotic solution is
stable, while source and saddle points provide unstable solutions. Because the exact cosmological solutions derived before \cite{dial} are related to the asymptotic solutions at the equilibrium points. From such analysis, we extract information related to the stability properties of the solutions derived
in \cite{dial}.

By definition, it follows that $y$ and $v$ are positively defined, that is,
$y\geq0$, $v\geq0$, while for $\omega<0$ and $\beta=-1$ from the algebraic
equation \eqref{dd.26}, it follows that $\left\vert \omega x^{2}\right\vert
\leq1$, $\left\vert \beta z^{2}\right\vert \leq1$, $0\leq y\leq1$ and $0\leq
v\leq1$. Otherwise, for other values of the free parameters $\omega,~\beta$
parameters $\left\{  x,y,z,v\right\}  $ are not constrained in the finite
regime and can take values at the infinity regime.
\section{Dynamical analysis at the finite regime}

\label{sec4}

\begin{figure}[ptb]
\centering\includegraphics[width=1\textwidth]{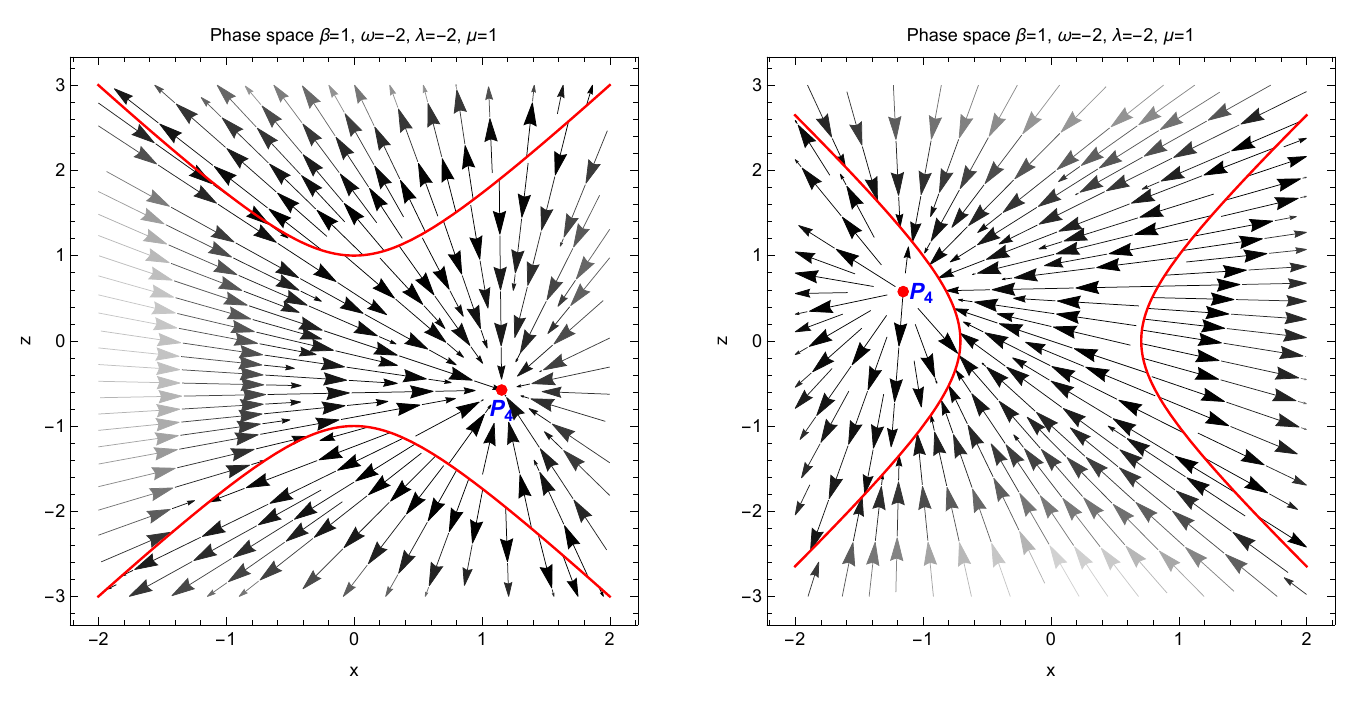}\caption{Two-dimensional
phase-space portrait of the dynamical system on the surface $y=0$. The left plot
is for $\beta=1$ and $\omega=-2$, while the right plot is for $\beta=-1$ and
$\omega=2$. With red lines are the family of points $P_{1}^{\pm}$. From the
phase-space portrait, points $P_{1}^{\pm}$ define surfaces where the
trajectories can move from the finite to the infinity regimes and vice versa.}%
\label{ap1}%
\end{figure}

The equilibrium points $P=\left(  x\left(  P\right), y\left(  P\right)
,z\left(  P\right)  \right)  $ of the field equations \eqref{eq-a}, \eqref{eq-b} and \eqref{eq-c} at the finite regime
are:
\[P_{1}^{\pm}=\left(  x_{1},0,\pm\sqrt{-\frac{1+\omega x_{1}^{2}}{\beta}}\right),\]
with $v\left(  P_{1}^{\pm}\right)  =0$, and $w_{eff}\left(  P_{1}^{\pm
}\right)  =1+\frac{8}{\sqrt{3}}x_{1}$. Because $y\left(  P_{1}^{\pm
}\right)  =0$\textbf{, }$v\left(  P_{1}^{\pm}\right)  =0$ we infer
that $\lambda\phi\left(  P_{1}^{\pm}\right)  \rightarrow-\infty$ and
$\mu\psi\left(  P_{1}^{\pm}\right)  \rightarrow-\infty$. $P_{1}^{\pm}$ are two sets of equilibrium points that exist for $\beta<0$ and $x_1=0$ or $\beta<0$ and $\frac{1}{x_1^2}+\omega \geq 0$ or $\beta >0$ and $\frac{1}{x_1^2}+\omega \leq 0$. As a particular case, we use
$\beta=\pm 1$, to simplify calculations. The asymptotic solutions at $P_{1}^{\pm}$ describe
accelerated universes for $x_{1}<-\frac{\sqrt{3}}{6}$. The eigenvalues of the
linearized system are derived $e_{1}\left(  P_{1}^{\pm}\right)  =0$,
$e_{2}\left(  P_{1}^{\pm}\right)  =3+\sqrt{3}\left(  4+\lambda\right)  x_{1}$
and $e_{3}\left(  P_{1}^{\pm}\right)  =6+8\sqrt{3}x_{1}\pm2\sqrt{3}\mu
\sqrt{-\frac{1+\omega x_{1}^{2}}{\beta}}$. Note that the eigenvector associated with $e_{1}\left(  P_{1}^{\pm}\right)=0$ is $\left(\frac{\sqrt{\beta  \left(-\omega  x_1^2-1\right)}}{x_1 \omega},0,1\right)$ is proportional to the normal vector to the sets given by $\left(1,0,\frac{x_1 \omega }{\sqrt{\beta  \left(-\omega x_1^2 -1\right)}}\right)$ this means that the sets are normally hyperbolic. Their stability is determined by the nonzero eigenvalues $e_2(P_1^{\pm}), e_3(P_1^{\pm})$. To understand the
stability properties of points $P_{1}^{\pm}$ in Fig.  \ref{ap1}, we plot the
two-dimensional phase-space portraits for the dynamical system on the surface
with $y=0$. From the phase-space portraits, we remark that points $P_{1}^{\pm
}$ define surfaces where the trajectories move from infinity to the finite
regimes and vice versa.

\[
P_{2}=\left(  \frac{4+\lambda}{\sqrt{3}\omega},\sqrt{\frac{\left(
4+\lambda\right)^2  +3\omega}{3\omega}},0\right), 
\]
with $v\left(  P_{2}\right)  =0$, and $w_{eff}\left(  P_{2}\right)
=-1-\frac{2\lambda\left(  4+\lambda\right)  }{\omega}$.

\begin{figure}[pbt]
    \centering
    \includegraphics[width=0.55\textwidth]{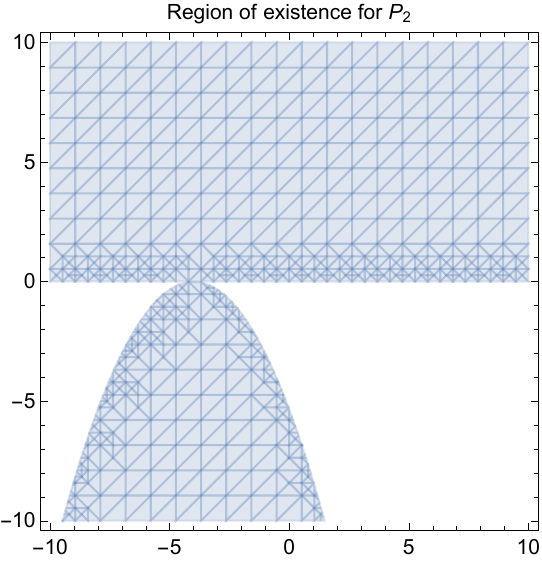}
\caption{Region of existence of point $P_2$ where $v(P_2)= 0.$} \label{exist-2}
\end{figure}

 Fig. \ref{exist-2} shows the region of existence of point $P_2$ where $v(P_2)= 0.$

\begin{figure}[ptb]
\centering
\includegraphics[width=0.6\textwidth]{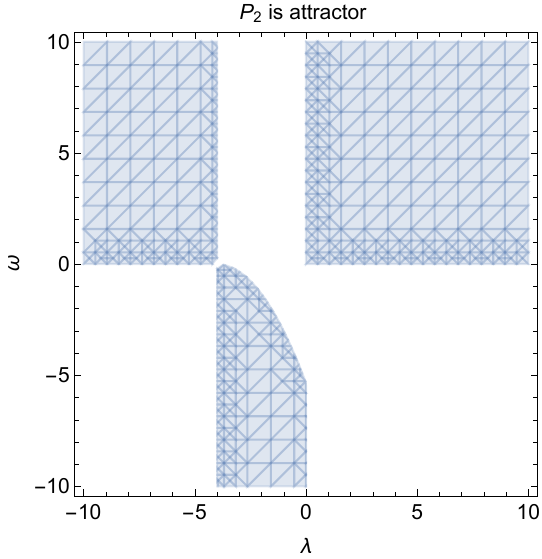}\caption{Region plot
on the space of the free parameters $\lambda$ and $\omega$~where point $P_{2}$
is an attractor. }%
\label{ap2}%
\end{figure}

For this point, only the potential of the second scalar field is zero, from where we infer
that $\mu\psi\left(  P_{2}\right)  \rightarrow-\infty$. \ The point
is real when $\frac{\left(  4+\lambda\right)^2  +3\omega}{3\omega}>0$. For $\beta=-1$ and $\omega<0$, it follows that the point is physically accepted for
$-\frac{\left\vert \omega\right\vert }{\sqrt{3}}<4+\lambda<\frac{\left\vert
\omega\right\vert }{\sqrt{3}}$. The eigenvalues are derived $e_{1}\left(
P_{2}\right)  =-3-\frac{\left(  4+\lambda\right)  ^{2}}{\omega},~e_{2}\left(
P_{2}\right)  =-3-\frac{\left(  4+\lambda\right)  ^{2}}{\omega}$ and
$e_{3}\left(  P_{2}\right)  =-\frac{2\lambda\left(  4+\lambda\right)  }%
{\omega}$. Hence, for $\beta=-1$, $\omega<0$ point $P_{2}$ is an attractor
when $\left\{  0\leq4+\lambda\leq\sqrt{3\left\vert \omega\right\vert }%
,-\frac{16}{3}<\omega\leq-1\right\}  $ or $\left\{  -4\leq\lambda\leq
0,~\omega\leq-\frac{16}{3}\right\}  $ or $\left\{  0\leq4+\lambda\leq\sqrt
{3}\left\vert \omega\right\vert \right\}  $. For other values of $\beta$ and $\omega$, point $P_{2}$ describes a stable solution when $\lambda=-4$ and
for a nonzero parameter $\omega$; or $\lambda=0$ and $\omega\leq-\frac{16}{3}$, or
$\omega>0$, $-4<\lambda<0$ and $\left(  4+\lambda\right)  ^{2}\leq-3\omega$
and $\omega>0$ with $\left\{  \lambda<-4\text{, }\lambda>0\right\}  $. In Fig. 
\ref{ap2} we give the region of the space of the free parameters $\lambda$ and
$\omega$ where point $P_{2}$ is an attractor. 

\begin{figure}[pbt]
    \centering
    \includegraphics[width=0.6\textwidth]{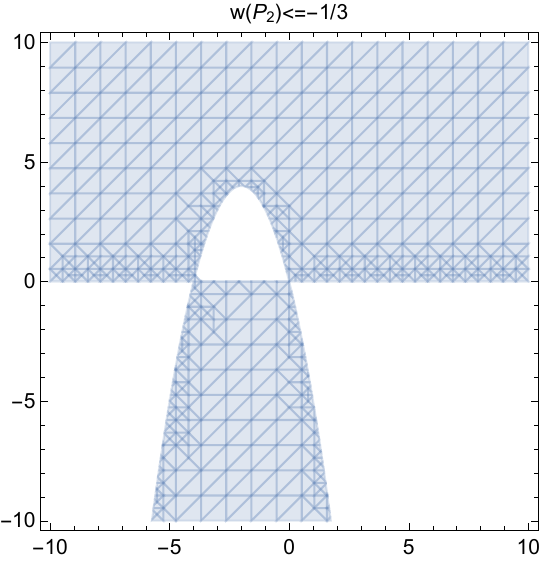}
\caption{Region for the free parameters $\lambda$ and $\omega$ where point $P_2$ describes an accelerated solution.} \label{acc-p2}
\end{figure}

Fig. \ref{acc-p2}  displays a region for the free parameters $\lambda$ and $\omega$ where point $P_2$ describes an accelerated solution.

\begin{figure}[ptb]
    \centering
    \includegraphics[scale=0.6]{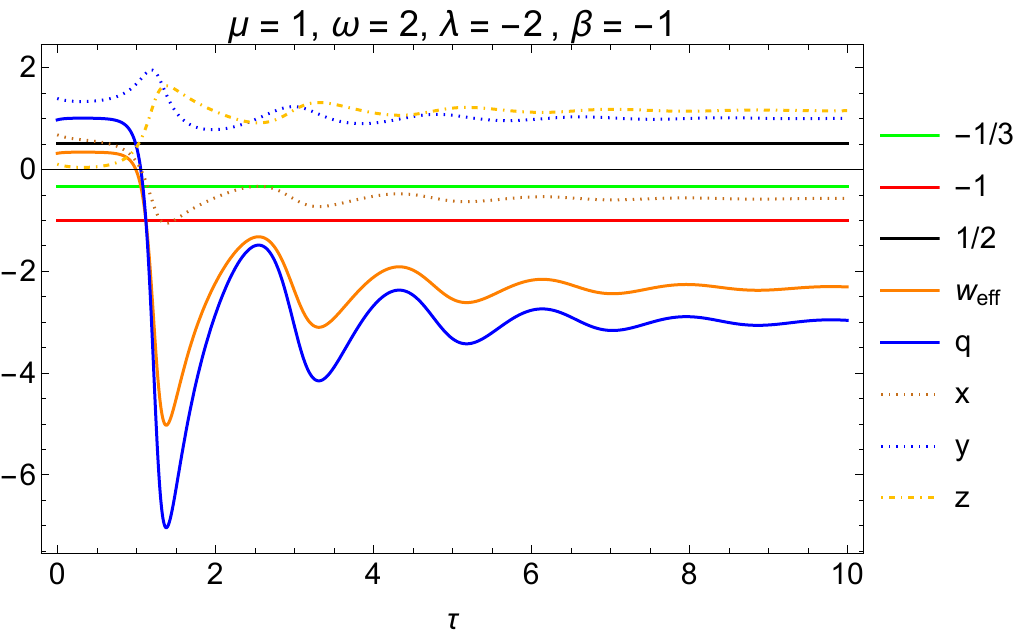}
    \includegraphics[scale=0.6]{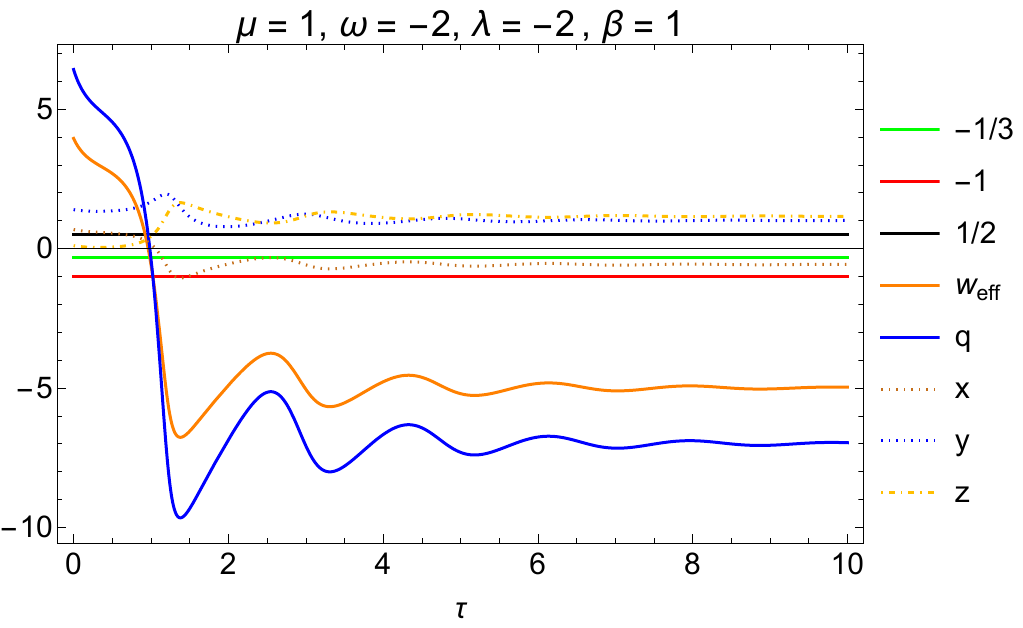}
    \caption{Evolution of $w_{eff}$ and $q$ evaluated at a numerical solution of system \eqref{eq-a}, \eqref{eq-b} and \eqref{eq-c} with initial conditions near $P_2$, i.e. $x(0)=\frac{\lambda +4}{\sqrt{3} \omega }+0.1,y(0)=\frac{\sqrt{(\lambda
   +4)^2+3 \omega }}{\sqrt{3} \sqrt{\omega }}+0.1,z(0)=0.1$ for different values of the parameters $\mu, \beta, \omega, \lambda.$ We see that for early time $w_{eff}>-\frac{1}{3}$ and $q>0$, meaning there is an early-time deceleration phase. Later, it enters a matter-dominated phase when $w_{eff}=0$ and $q=\frac{1}{2}$ but quickly after that, we see that the behaviour describes late time acceleration for $w_{eff}<-\frac{1}{3}$ and $q<0.$ We also see a transient de Sitter behaviour since both cross the value $w_{eff}=q=-1.$ Finally, they cross to the phantom regime $w_{eff}<-1$ and $q<-1$ with a transient damped oscillation in both parameters.}
    \label{weff-1}
\end{figure}
\begin{figure}[ptb]
\centering\includegraphics[width=1\textwidth]{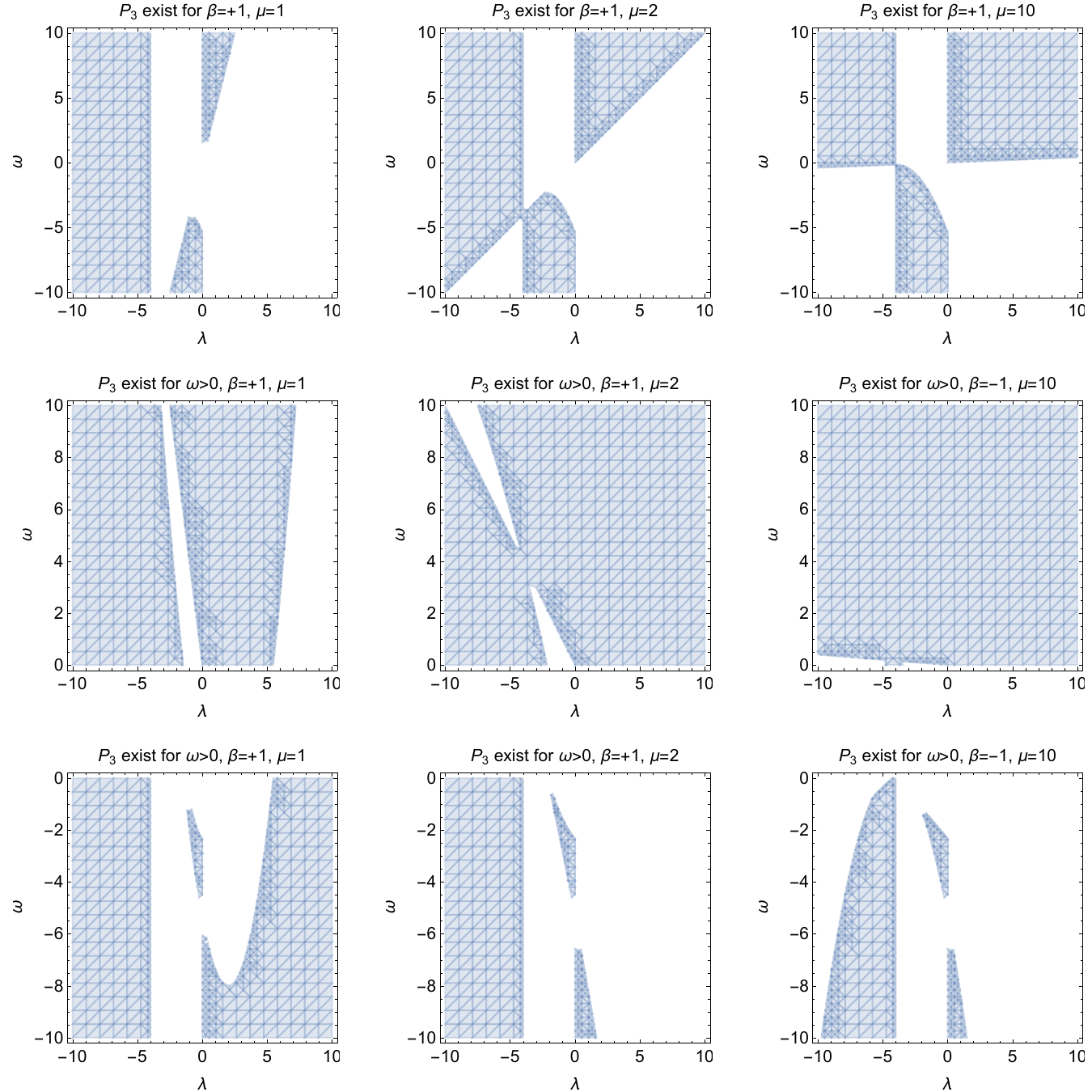}\caption{Region plot on
the space of the free parameters $\lambda$ and $\omega$ for different values
of $\mu$ where point $P_{3}$ exists. The figures in the first row are for
$\beta=+1$; while figures of the second row and the third row are for
$\beta=-1.$}%
\label{ap3a}%
\end{figure}
\begin{figure}[ptb]
\centering\includegraphics[width=1\textwidth]{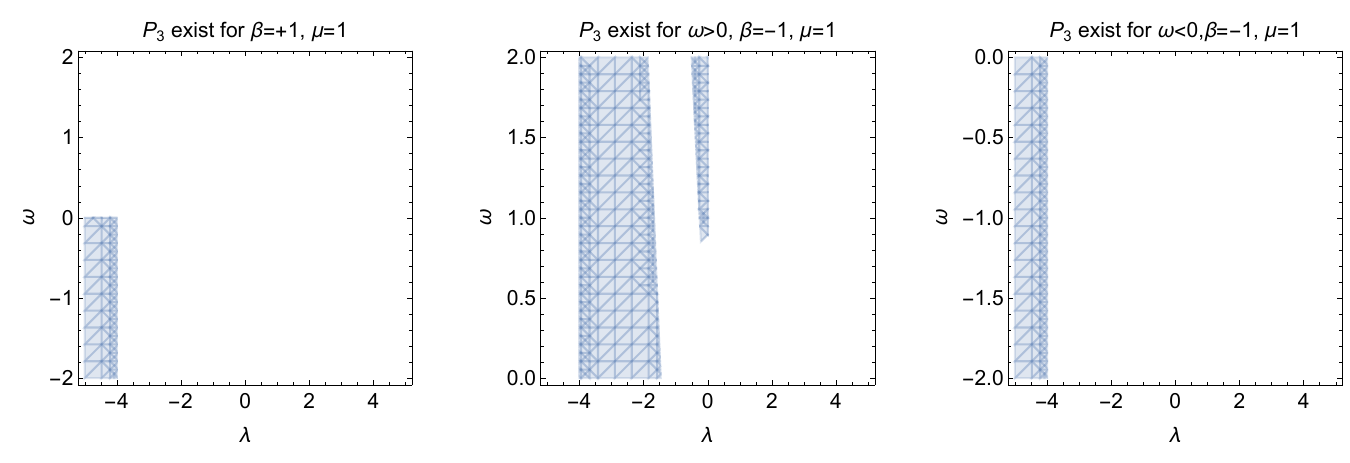}\caption{Region plot on
the space of the free parameters $\lambda$ and $\omega$ for $\mu$ where point
$P_{3}$ is an attractor. }%
\label{ap3b}%
\end{figure}
\begin{figure}[ptb]
\centering\includegraphics[width=1\textwidth]{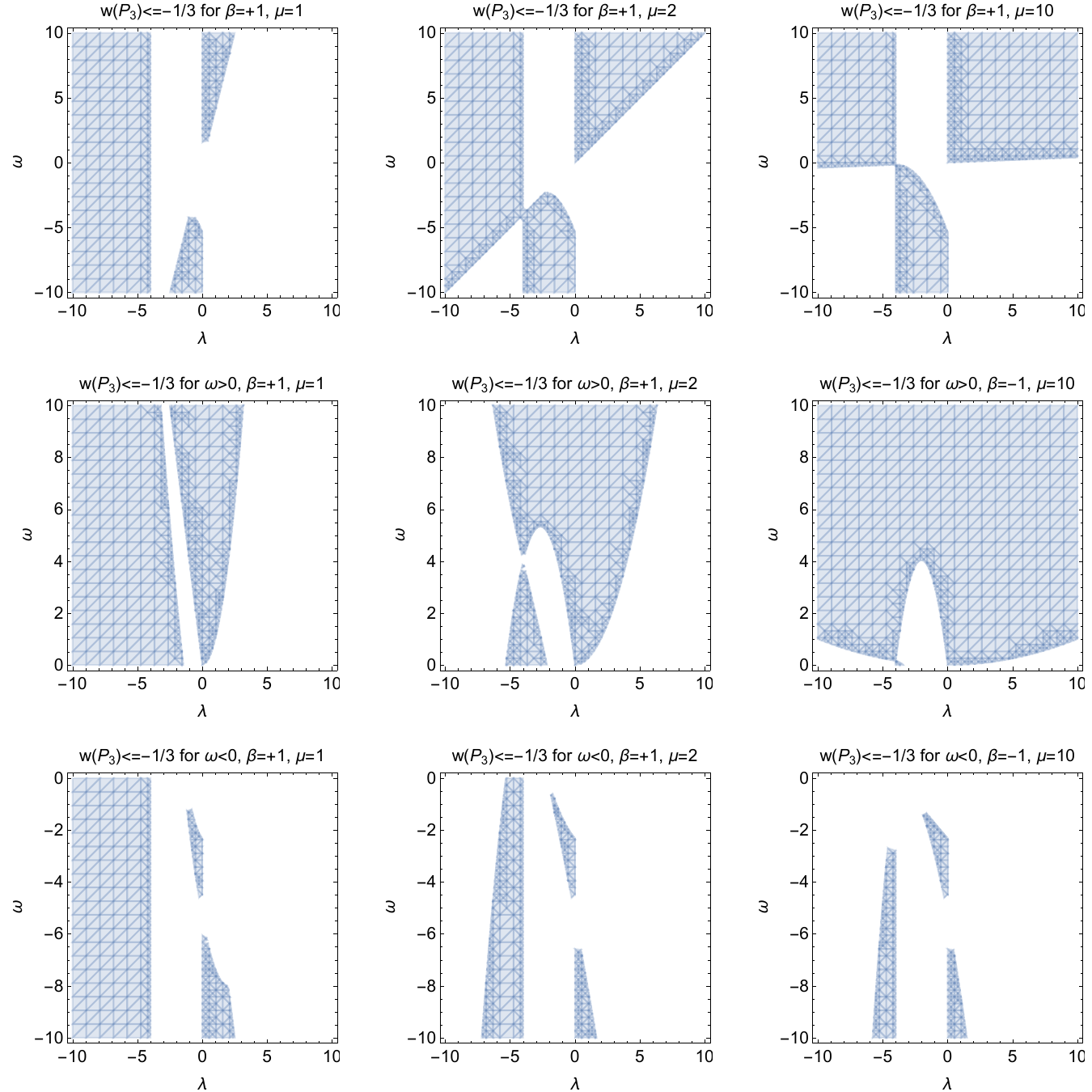}\caption{Region plot on
the space of the free parameters $\lambda$ and $\omega$ for different values
of $\mu$ where $w\left(  P_{3}\right)  < -\frac{1}{3}$. Figures of the first
row are for $\beta=+1$, while figures of the second row and the third row are
for $\beta=-1.$}%
\label{ap3c}%
\end{figure}

In Fig. \ref{weff-1}, the evolution of $w_{eff}$ and $q$ is presented. The parameters are evaluated at a numerical solution of \eqref{eq-a}, \eqref{eq-b} and \eqref{eq-c} with initial conditions near $P_2$, i.e. $x(0)=\frac{\lambda +4}{\sqrt{3} \omega }+0.1,y(0)=\frac{\sqrt{(\lambda
   +4)^2+3 \omega }}{\sqrt{3} \sqrt{\omega }}+0.1,z(0)=0.1$. This figure shows that for early time $w_{eff}>-\frac{1}{3}$ and $q>0$; therefore, there is an early-time deceleration phase. Later evolution enters a matter-dominated phase when $w_{eff}=0$ and $q=\frac{1}{2}$. Quickly after that, we see that the behaviour describes late time acceleration for $w_{eff}<-\frac{1}{3}$ and $q<0.$ We also see a transient de Sitter behaviour since both cross the value $w_{eff}=q=-1.$ Finally, they cross to the phantom regime $w_{eff}<-1$ and $q<-1$ with a transient damped oscillation in both parameters. The evolution of the $x, y$ and $z$ variables in the time $\tau$ are presented for completeness.

\[
P_{3}=\left(  \frac{\left(  4+\lambda\right)  \mu^{2}}{\sqrt{3}\left(
\beta\lambda^{2}+\omega\mu^{2}\right)  },\sqrt{-\frac{\left(  4\beta
\lambda-\omega\mu^{2}\right)  \left(  3\left(  \beta\lambda^{2}+\omega\mu
^{2}\right)  +\mu^{2}\left(  4+\lambda\right)  ^{2}\right)  }{3\left(
\beta\lambda^{2}+\omega\mu^{2}\right)  ^{2}}},\frac{\left(  4+\lambda\right)
\lambda\mu}{\sqrt{3}\left(  \beta\lambda^{2}+\omega\mu^{2}\right)  }\right), 
\]
with $v\left(  P_{3}\right)  =\sqrt{\frac{\beta\lambda\left(  4+\lambda
\right)  \left(  3\beta\lambda^{2}+\mu^{2}\left(  \left(  4+\lambda\right)
^{2}+3\omega\right)  \right)  }{3\left(  \beta\lambda^{2}+\omega\mu
^{2}\right)  ^{2}}}$, and $w_{eff}\left(  P_{3}\right)  =-1-\frac
{2\lambda\left(  4+\lambda\right)  \mu^{2}}{3\left(  \beta\lambda^{2}%
+\omega\mu^{2}\right)  }$. Therefore $\phi\left(  P_{3}\right)
$ and $\psi\left(  P_{4}\right)  $ are defined in the finite regime. In Fig.  \ref{ap3a}, we present the region in the space of the free
parameters $\lambda$ and $\omega$ where the point is real and physically accepted. Because of the nonlinear form of the eigenvalues, the stability properties are investigated numerically. In Fig.  \ref{ap3b} we present the
region in the space of the free parameters $\lambda$ and $\omega$ where the equilibrium point $P_{3}$ is an attractor for $\mu=1$. Moreover, in Fig. 
\ref{ap3c} we present the values of the free parameters $\lambda$ and $\omega$ where the asymptotic solution at point $P_{3}$ describes an accelerated universe, i.e. $w\left(  P_{3}\right)  <-\frac{1}{3}$.

\begin{figure}[pbt]
    \centering
    \includegraphics[scale=1.0]{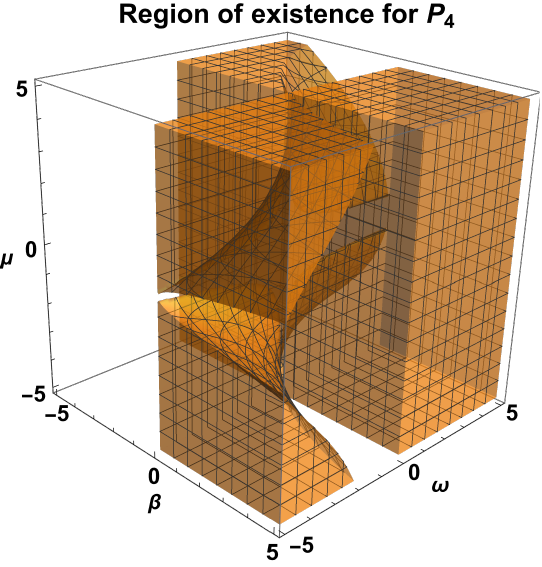}
    \caption{Region of existence of point $P_4$ where $v(P_4)\geq 0$, that is $\frac{\mu ^2}{3 \beta }+\frac{16}{3 \omega }+1\geq 0$.}
    \label{exist-1}
\end{figure}
\[
P_{4}=\left(  \frac{4\sqrt{3}}{3\omega},0,\frac{\sqrt{3}\mu}{3\beta}\right), 
\]
with $v\left(  P_{4}\right)  =\sqrt{1+\frac{1}{3}\left(  \frac{\mu^{2}}{\beta
}+\frac{16}{\omega}\right)  }$, and $w_{eff}\left(  P_{4}\right)
=-1-\frac{3\mu^{2}}{2\beta}$. Hence, $\lambda\phi\left(
P_{4}\right)  \rightarrow-\infty$, and $\psi\left(  P_{4}\right)
$ is defined in the finite regime. The point is real and physically
accepted for the regions shown in Fig.   \ref{exist-1}. Some particular cases of interest are the following:  for $-\frac{3}{16}-\frac{\mu^{2}}{16\beta}<\frac{1}{\omega}$; on the other hand for $\omega<0$ and $\beta=-1$, $P_{4}$ is physically accepted for $\mu^{2}<3$, $\left(  \mu^{2}-3\right)  \omega\geq16$. 

For $\beta=-1$, the asymptotic solution describes an accelerated universe for $\left\vert \mu\right\vert <\frac{2}{3}$, while for $\beta=+1$, the asymptotic solution always describes acceleration with $w_{eff}\left(  P_{4}\right)
|_{\beta=1}<-1$. The eigenvalues of the linearized system near the stationary point are $e_{1}\left(  P_{4}\right)  =-\frac{\mu^{2}}{\beta}+\frac{4\lambda
}{\omega}$, $e_{2}\left(  P_{4}\right)  =-3-\frac{\mu^{2}}{\beta}-\frac
{16}{\omega}$ and $e_{3}\left(  P_{4}\right)  =e_2\left(  P_{4}\right) .$

Consequently, the equilibrium point $P_{4}$ is an attractor for $\mu=0$~with
$\left\{  \lambda<0,\omega>0\right\}  $ or $\left\{  \lambda>0,\omega
<-\frac{16}{3}\right\}  $ for arbitrary $\beta$. In the case where $\beta=+1$,
point $P_{4}$ is an attractor for $\left\{  \lambda<-\frac{4\mu^{2}}{3+\mu
^{2}},~0<\omega<\frac{4\lambda}{\mu^{2}}\right\}  $, or $\left\{  -\frac
{4\mu^{2}}{3+\mu^{2}}<\lambda<0,~0<\omega\right\}  ,~$or $\left\{
-\frac{4\mu^{2}}{3+\mu^{2}}<\lambda<0,~\omega<-\frac{16}{3+\mu^{2}}\right\}  $ or $\left\{  \lambda>0,~\omega<-\frac{16}{3+\mu^{2}}\right\}  $ or$~\left\{
\lambda>0,~\frac{4\lambda}{\mu^{2}}<\omega\right\}  $. 

On the other hand, for $\beta=-1$, point $P_{4}$ is an attractor for $\mu^{2}<3$ with $\left\{
\lambda<0,\omega>0,\omega<-\frac{4\lambda}{\mu}\right\}  $ or $\left\{
\omega>-\frac{4\lambda}{\mu^{2}},\frac{\mu^{2}\left(  4+\lambda\right)  }%
{3}<\lambda,\omega<\frac{16}{\mu^{2}-3}\right\}  $, or $\mu>3$ with $\left\{
\lambda\leq-\frac{4\mu^{2}}{\mu^{2}-3},~0<\omega<\frac{16}{\mu^{2}-3}\right\}
$ or $\left\{  -\frac{4\mu^{2}}{\mu^{2}-3}<\lambda<0\text{,~}0<\omega
<-\frac{4\lambda}{\mu}\right\}  $ and for $\mu=3$, $\left\{  0<\omega
<-\frac{4\lambda}{3},~\lambda>0\right\}  $. 

\begin{figure}[ptb]
\centering\includegraphics[width=1\textwidth]{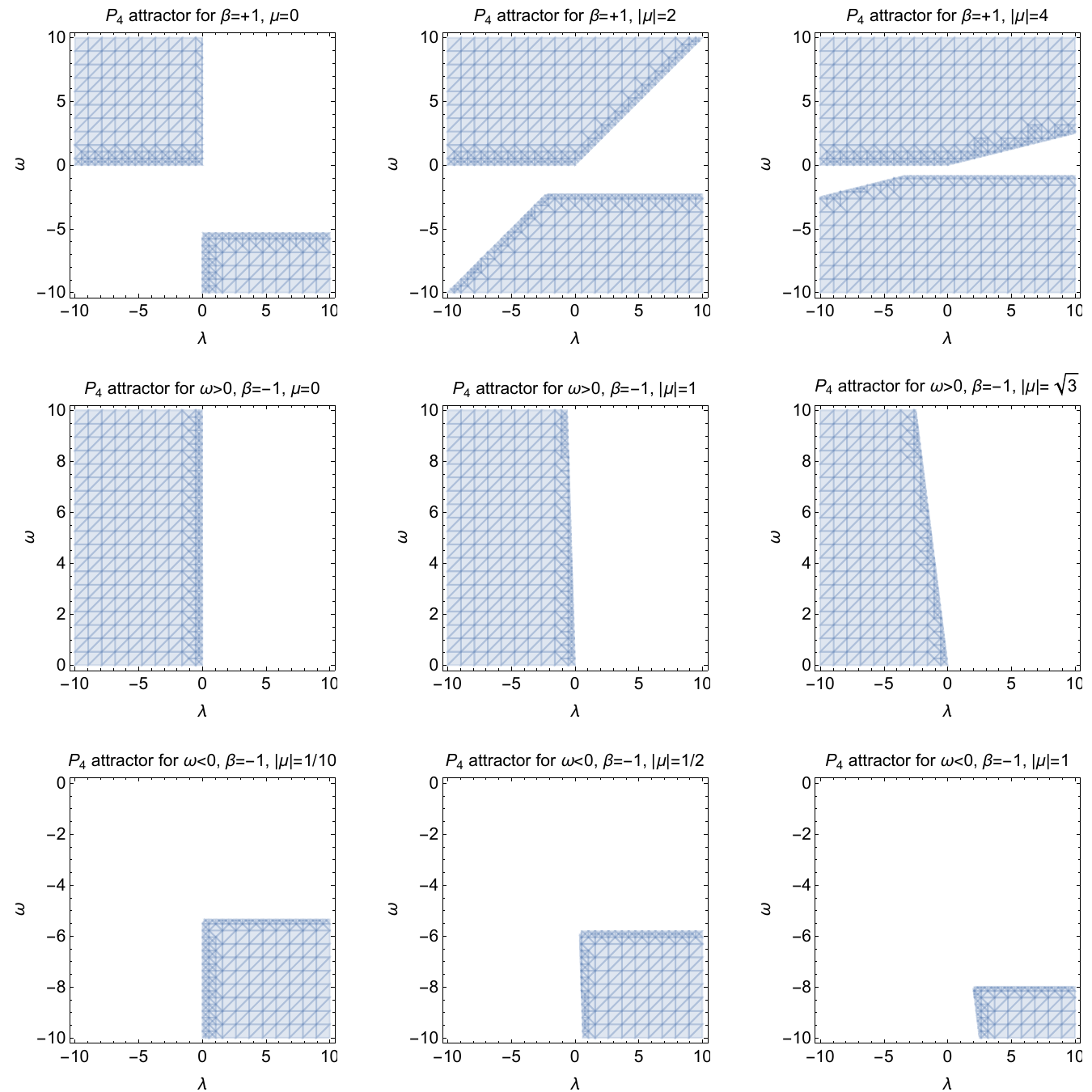}\caption{Region plot on
the space of the free parameters $\lambda$ and $\omega$ for different values
of $\mu^{2}$ where point $P_{4}$ is an attractor. The figures in the first row
are for $\beta=+1$; while figures of the second row are for $\beta=-1.$}%
\label{ap4}%
\end{figure}

Last but not least, for $\omega<0$,
$\beta=-1$ the equilibrium point is an attractor when $\mu^{2}<3,~\left(
\mu^{2}-3\right)  \omega\geq16$ and $4\lambda+\mu\omega>0$. In Fig.   \ref{ap4}
we present the graphic illustration of the latter regions.

In Table \ref{tab1}, we summarize the results of the dynamical analysis at the
finite regime.%

\begin{table}[tbp] \centering
\caption{Equilibrium points for the multi-torsion cosmological model \eqref{eq-a}, \eqref{eq-b} and \eqref{eq-c} in the finite regime.}%
\begin{tabular}
[c]{ccccc}\hline\hline
\textbf{Points} & $\mathbf{w}_{eff}$& $q$ & \textbf{Acceleration?} &
\textbf{Attractor?}\\\hline
$P_{1}^{\pm}$ & $1+\frac{8}{\sqrt{3}}x_{1}$ & $4 \sqrt{3} x_1+2$ & $x_{1}<-\frac{\sqrt{3}}{6}$ &
No\\
$P_{2}$ & $-1-\frac{2\lambda\left(  4+\lambda\right)  }{\omega}$& $-\frac{\lambda  (\lambda +4)+\omega }{\omega }$& Yes (Fig.  \ref{acc-p2}) & Yes
(Fig.   \ref{ap2})\\
$P_{3}$ & $-1-\frac{2\lambda\left(  4+\lambda\right)  \mu^{2}}{3\left(
\beta\lambda^{2}+\omega\mu^{2}\right)  }$ & $-\frac{\beta  \lambda ^2+\mu ^2 (\lambda  (\lambda +4)+\omega )}{\beta 
   \lambda ^2+\mu ^2 \omega }$& Yes (Fig.  \ref{ap3c}) & Yes (Fig. 
\ref{ap3b})\\
$P_{4}$ & $-1-\frac{2\mu^{3}}{2\beta}$& $-\frac{\beta +\mu ^2}{\beta }$ & $\beta +\mu ^2<0$ & \\  &&&  or $\beta >0$ & Yes (Fig.  \ref{ap4}%
)\\\hline\hline
\end{tabular}
\label{tab1}%
\end{table}%

\begin{figure}[ptb]
    \centering
    \includegraphics[scale=0.7]{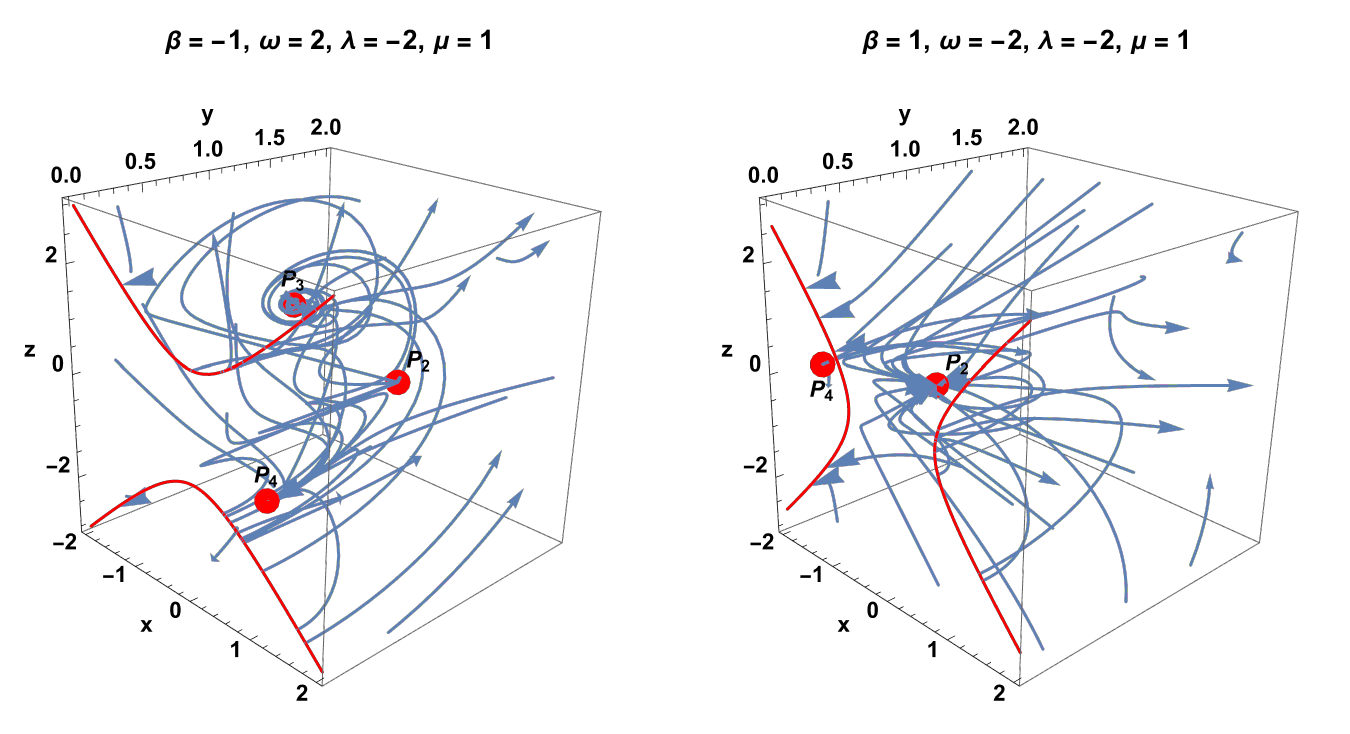}
    \caption{Three-dimensional phase-space portrait of the full system \eqref{eq-a}, \eqref{eq-b} and \eqref{eq-c}. In the left plot, the values $\lambda=-2, \mu=1, \beta=-1, \omega=2$ were considered. In the right plot, the values $\lambda=-2, \mu=1, \beta=1, \omega=-2$ were considered. The planes that contain the hyperbolas in both portraits are the $y=0$ projections shown in Fig.  \ref{ap1}}
    \label{Stream3d}
\end{figure}

To finish this section, in Fig. \ref{Stream3d} is shown a three-dimensional phase-space portrait of the full system \eqref{eq-a}, \eqref{eq-b} and \eqref{eq-c}. The values $\lambda=-2, \mu=1, \beta=-1, \omega=2$ were considered in the left plot. The values $\lambda=-2, \mu=1, \beta=1, \omega=-2$ were considered in the right plot. The planes that contain the hyperbolas in both portraits are the $y=0$ projections shown in Fig.  \ref{ap1}.

\section{Dynamical analysis at the infinity regime}

\label{sec5}

Owing to the fact that the dynamical system \eqref{eq-a}, \eqref{eq-b} and \eqref{eq-c} is
non-compact, there could be features in the asymptotic regime
which are non-trivial for the global dynamics. Thus, we will extend our
study using the Poincar\'e method to complete the phase space analysis. We define the Poincar\'e variables
\[x=\frac{\rho}{\sqrt{1-\rho^{2}}}\sin\theta_{1}\cos\theta_{2}, \quad y=\frac{\rho}{\sqrt
{1-\rho^{2}}} \sin \theta_1 \sin\theta_{2}, \quad z=\frac{\rho}{\sqrt{1-\rho^{2}}}\cos\theta_{1}.
\]
The inverse transformation is  
\[\rho= \frac{\sqrt{x^2+y^2+z^2}}{\sqrt{x^2+y^2+z^2+1}}, \quad \theta_1= \tan^{-1}\left(\frac{\sqrt{x^2+y^2}}{z}\right), \quad \theta_2= \tan^{-1}\left(\frac{y}{x}\right).\]

We have chosen these spherical coordinates where $\theta_1$ is the azimuth angle measured with respect to the $z$-axis. Because $y$ is non-negative by definition, and $x$ and $z$ can have either sign ($\theta_2$ is the  polar angle) the physical region is $0\leq\rho\leq1$, $\theta_{1}\in\left[0,\pi\right]  $ and $\theta
_{2}\in\lbrack 0, \pi].$ Infinity is reached when $\rho\rightarrow1$. 

We consider the new independent variable $\frac{d}{dT}=\sqrt{1-\rho^{2}}\frac{d}{d\tau},$ thus the
dynamical system \eqref{dd.20}-\eqref{dd.22} with the use of the constraint
equation \eqref{dd.26} is expressed in terms of the variables $\left\{
T,\rho,\theta_{1},\theta_{2}\right\}  $ in the following form%
\begin{equation}
\frac{d\rho}{dT}=f_{1}\left(  \rho,\theta_{1},\theta_{2}\right), 
\end{equation}%
\begin{equation}
\frac{d\theta_{1}}{dT}=f_{2}\left(  \rho,\theta_{1},\theta_{2}\right), 
\end{equation}%
\begin{equation}
\frac{d\theta_{2}}{dT}=f_{3}\left(  \rho,\theta_{1},\theta_{2}\right), 
\end{equation}
where $f_{1},~f_{2}$ and $f_{3}$ are nonlinear expressions. At the finite
regime, $\rho\rightarrow1$, the leading terms of the latter dynamical system reads%
\begin{small}
\begin{align}
\frac{1}{\sqrt{1-\rho}}\frac{d\rho}{dT} & = -\frac{3}{4} \left((2 \beta -\omega ) \cos \left(2 \theta _1\right)+2 \beta +2
   \omega  \sin ^2\left(\theta _1\right) \cos \left(2 \theta _2\right)+\omega
   \right), \label{eqrho}
\end{align}
\begin{align}
\frac{d\theta_{1}}{dT}  &  =\frac{\sqrt{3}}{8 \beta \omega }  \Big[\beta  \cos \left(\theta _1\right) \Big(\cos
   \left(\theta _2\right) \left(\cos \left(2 \theta _1\right) (16 \beta -\lambda 
   (\omega +1)-16 \omega )+16 \beta +\lambda  \omega +\lambda +16 \omega \right) \nonumber \\
   & -2 \lambda  (\omega +1) \sin ^2\left(\theta _1\right) \cos \left(3 \theta_2\right)\Big) \nonumber \\
   & -\mu  \omega  \left((2 \beta +3 \omega -3) \sin \left(\theta
   _1\right)+(2 \beta -\omega +1) \sin \left(3 \theta _1\right)+4 (\omega +1) \sin
   ^3\left(\theta _1\right) \cos \left(2 \theta _2\right)\right)\Big],
\end{align}
\begin{align}
\frac{d\theta_{2}}{dT} & =\frac{\sqrt{3} \sin \left(\theta _1\right) \sin \left(\theta _2\right)
   \left(-4 \beta  \cot ^2\left(\theta _1\right)+\lambda  \omega  \cos ^2\left(\theta
   _2\right)-\lambda  \sin ^2\left(\theta _2\right)\right)}{\omega}.
\end{align}   
\end{small}

We represent the infinite regime in the right hemisphere  of the Poincaré sphere, 
\[X= \rho \sin\theta_{1}\cos\theta_{2}, \quad Y= \rho \sin \theta_1 \sin\theta_{2}, \quad Z= \rho \cos\theta_{1}.
\]

The radial equations do not contain the radial coordinate, so the equilibrium points can only be obtained using the angular equations.
Setting $\frac{d\theta_{1}}{dT}=0, \frac{d\theta_{2}}{dT}=0$, we obtain the equilibrium points listed in table \ref{TII}. These points' stability is studied by first analyzing the angular coordinates' stability and then deducing the stability in the radial direction from the equation \eqref{eqrho}. In these variables, infinity is reached when $\rho\rightarrow1$. Hence, a point at infinity is stable in the radial direction when $\frac{d\rho(T)}{dT}>0$, which means that $\rho(T)$ is a monotonic increasing function that tends towards $\rho=1$ from below.  When $\frac{d\theta_{1}}{dT}<0,$ the equilibrium points at infinity are either saddles or sources, depending on the radial stability. When $\frac{d\rho(T)}{dT}=0$, the method is not conclusive, so we resort to numerical inspection.

In the following analysis, the eigenvalue $e_1(Q)$ corresponds to the radial direction the equation \eqref{eqrho} gives. Consequently, the equilibrium points $Q:=(X,Y,Z)$ at the infinite regime are
$Q_{1}^{+}= \left(\frac{1}{\sqrt{\omega +1}},\sqrt{\frac{\omega }{\omega +1}},0\right),$ and $Q_{2}^{+}= \left(-\frac{1}{\sqrt{\omega +1}},\sqrt{\frac{\omega }{\omega +1}},0\right),$ they exist for $\omega\geq 0.$

The eigenvalues are $e_1(Q_{1}^{+})=e_1(Q_{2}^{+})=-\frac{3 \omega }{\omega +1},$ $e_2(Q_{1}^{+})=-\frac{2 \sqrt{3} \lambda }{\sqrt{\omega +1}}=-e_2(Q_{2}^{+}), e_3(Q_{1}^{+})=-\frac{\sqrt{3} (\lambda +4)}{\sqrt{\omega +1}}=-e_3(Q_{2}^{+}).$ This means that $Q_1^+$ is source for  $\{\omega>0, \lambda<-4\}$ and a saddle for $\{-4<\lambda <0$, $ \omega >0\}$.

On the other hand, $Q_2^+$ is a source for 
$\{\lambda >0$, $ \omega >0\}$ and a saddle for $\{-4<\lambda <0$, $ \omega >0\}.$

The following points are 
$Q_{3}^{+}= \left(\sqrt{\frac{\beta}{\beta-\omega}},0,\sqrt{\frac{\omega}{\omega-\beta}}\right),$
$Q_{4}^{+}= \left(-\sqrt{\frac{\beta}{\beta-\omega}},0,-\sqrt{\frac{\omega}{\omega-\beta}}\right),$
$Q_{5}^{+}= \left(\sqrt{\frac{\beta}{\beta-\omega}},0,-\sqrt{\frac{\omega}{\omega-\beta}}\right),$
$Q_{6}^{+}= \left(-\sqrt{\frac{\beta}{\beta-\omega}},0,\sqrt{\frac{\omega}{\omega-\beta}}\right),$ they all exist for $\omega <0$ and $\beta\geq 0$ or $\omega >0$ and $\beta\leq 0$ or $\omega=0$ and $\beta \neq 0.$  Also, the eigenvalue in the radial direction $e_1(Q_i^+)=0$ for $i=3,4,5,6.$

As a particular case, we use $\beta=\pm 1$ to simplify calculations.

The eigenvalues for $Q_3^+$ are $e_2(Q_3^+)=\frac{\sqrt{3} \sqrt{\beta } (\lambda +4)}{\sqrt{\beta -\omega }},e_3(Q_3^+)=\frac{2 \sqrt{3} \left(\mu  \sqrt{\omega }+4 i \sqrt{\beta }\right)}{\sqrt{\omega -\beta }}.$

In the 2-dimensional projection, $Q_3^+$ is an attractor for $\{\lambda<-4, \mu<0, \omega<-\frac{16}{\mu^2}\}.$

$Q_3^+$ is a source for $\{\omega=0, \lambda>-4$ or $\lambda>-4, \mu <0, -\frac{16}{\mu ^2}<\omega <0\}$ or $\{\lambda>-4, \mu \geq 0, \omega <0\}$ it is also a saddle in any other case.

The eigenvalues for the points $Q_4^+$  are the opposite of the eigenvalues of $Q_3^+,$ that is $e_2(Q_4^+)=-\frac{\sqrt{3} \sqrt{\beta } (\lambda +4)}{\sqrt{\beta -\omega }}=-e_2(Q_3^+), e_3(Q_4^+)=\frac{2 \sqrt{3} \left(-\mu  \sqrt{\omega }-4 i \sqrt{\beta
   }\right)}{\sqrt{\omega -\beta }}=-e_3(Q_3^+)$. That means that the dynamics of $Q_4^+$ are reversed. 

Similarly, for $Q_5^+$ and $Q_6^+$ their respective eigenvalues are opposite, they are  $e_2(Q_5^+)=\frac{\sqrt{3} \sqrt{\beta } (\lambda +4)}{\sqrt{\beta -\omega }}, e_3(Q_5^+)=\frac{2 \sqrt{3} \left(-\mu  \sqrt{\omega }+4 i \sqrt{\beta }\right)}{\sqrt{\omega
   -\beta }}$ and  $e_2(Q_6^+)=-\frac{\sqrt{3} \sqrt{\beta } (\lambda +4)}{\sqrt{\beta -\omega }}, e_3(Q_6^+)=\frac{2 \sqrt{3} \left(\mu  \sqrt{\omega }-4 i \sqrt{\beta }\right)}{\sqrt{\omega
   -\beta }},$ since the dynamics are reversed, we present the analysis for $Q_5^+.$ 
  
   In the 2-dimensional projection, this point is an attractor for $\{\lambda <-4, \mu >0, \omega <-\frac{16}{\mu ^2}\},$ and a source for $\{\omega=0, \lambda>-4$ or $\lambda >-4, \mu \leq 0, \omega <0\}$ or $\{\lambda >-4, \mu >0, -\frac{16}{\mu ^2}<\omega <0\}$  it is also a saddle in any other case.

   The points $Q_{7}^{+}=\left(\frac{4 \beta }{\sqrt{16 \beta ^2+\mu ^2 \omega ^2}},0,\frac{\mu  \omega }{\sqrt{16 \beta ^2+\mu ^2 \omega
   ^2}}\right),$ and
$Q_{8}^{+}=\left(-\frac{4 \beta }{\sqrt{16 \beta ^2+\mu ^2 \omega ^2}},0,-\frac{\mu  \omega }{\sqrt{16 \beta ^2+\mu ^2 \omega
   ^2}}\right),$ both exist for $16 \beta ^2+\mu ^2 \omega ^2>0,$ share the same eigenvalue in the radial direction $e_1(Q_{7}^{+})=e_1(Q_{8}^{+})=-\frac{3 \omega  \left(\mu ^2 \omega +16\right)}{\mu ^2 \omega ^2+16}$ and also have opposite eigenvalues in the 2-dimensional projection, they are $e_2(Q_{7}^{+})=\frac{\sqrt{3} \left(4 \lambda -\mu ^2 \omega \right)}{\sqrt{\mu ^2 \omega ^2+16}}=-e_2(Q_{8}^{+}),  e_3(Q_{7}^{+})=-\frac{\sqrt{3} \left(\mu ^2 \omega +16\right)}{\sqrt{\mu ^2 \omega ^2+16}}=-e_3(Q_{8}^{+}).$  
 
   The point $Q_7^+$ is an attractor for $\{\beta >0,  \omega <0, -4
   \sqrt{-\frac{\beta }{\omega }}<\mu <4
   \sqrt{-\frac{\beta }{\omega }}, \lambda
   <\frac{\mu ^2 \omega }{4 \beta }\}$ or $\{\beta <0, \omega >0,
   \mu <-4 \sqrt{-\frac{\beta
   }{\omega }}, \lambda >\frac{\mu ^2 \omega
   }{4 \beta }\}$ or $\{ \mu >4
   \sqrt{-\frac{\beta }{\omega }}, \lambda
   >\frac{\mu ^2 \omega }{4 \beta
   }\},$ a source for $\{\beta >0, \omega <0,
   \mu <-4 \sqrt{-\frac{\beta
   }{\omega }}, \lambda >\frac{\mu ^2 \omega
   }{4 \beta }\}$ or $\{ \mu >4
   \sqrt{-\frac{\beta }{\omega }}, \lambda
   >\frac{\mu ^2 \omega }{4 \beta
   }\}$ or $\{\beta
   <0, \omega >0, -4 \sqrt{-\frac{\beta
   }{\omega }}<\mu <4 \sqrt{-\frac{\beta
   }{\omega }}, \lambda <\frac{\mu ^2 \omega
   }{4 \beta }\}$ and a saddle in any other case.

   The point $Q_8^+$ is an attractor for $\{\lambda <0, \omega <0, \beta <0\}$ or $
   \{\mu \neq 0, \lambda =0, \omega
   <0, \beta <0\}$ or $\{ \mu \neq 0,
   \lambda >0, \omega <0, \frac{\mu ^2
   \omega }{4 \lambda }<\beta <0\}$ a source for $\{\lambda <0, \omega >0, \beta >0\}$ or $
   \{\mu \neq 0, \lambda =0, \omega
   >0, \beta >0\}$ or $ \{\mu \neq 0,
   \lambda >0, \omega >0, 0<\beta
   <\frac{\mu ^2 \omega }{4 \lambda }\}$ and a saddle in any other case.

 Finally the points 
\[Q_{9}^{+}=\left(\frac{\mu }{\sqrt{-4 \beta  \lambda +\lambda ^2+\mu ^2 (\omega +1)}},\frac{\sqrt{4 \beta  \lambda -\mu ^2
   \omega }}{\sqrt{4 \beta  \lambda -\lambda ^2-\mu ^2 (\omega +1)}},\frac{\lambda }{\sqrt{-4 \beta  \lambda +\lambda
   ^2+\mu ^2 (\omega +1)}}\right)
,\]
\[Q_{10}^{+}=\left(-\frac{\mu }{\sqrt{-4 \beta  \lambda +\lambda ^2+\mu ^2 (\omega +1)}},\frac{\sqrt{4 \beta  \lambda -\mu ^2
   \omega }}{\sqrt{4 \beta  \lambda -\lambda ^2-\mu ^2 (\omega +1)}},-\frac{\lambda }{\sqrt{-4 \beta  \lambda +\lambda
   ^2+\mu ^2 (\omega +1)}}\right)
,\]
They exist for $\lambda <0,  \beta \geq \frac{\mu ^2 \omega
   }{4 \lambda }$ or $\mu \neq 0,  \lambda =0,  \omega \geq 0$ or $\lambda >0,  \beta \leq \frac{\mu ^2 \omega
   }{4 \lambda }.$ The eigenvalue in the radial direction for both points is $e_1(Q_{9}^{+})=e_1(Q_{10}^{+})=-\frac{3 \left(\beta  \lambda ^2+\mu ^2 \omega
   \right)}{-4 \beta  \lambda +\lambda ^2+\mu ^2
   (\omega +1)}.$ However the other two eigenvalues are complicated expressions that depend on the four parameters $\lambda, \mu, \omega, \beta,$ that is $e_2(Q_{9}^{+})=f_1(\lambda, \mu, \omega, \beta),$ $e_3(Q_{9}^{+})=f_2(\lambda, \mu, \omega, \beta)$ and $e_2(Q_{10}^{+})=g_1(\lambda, \mu, \omega, \beta),$ $e_3(Q_{10}^{+})=g_2(\lambda, \mu, \omega, \beta).$

   For some values in which these points exist, the real part of the eigenvalues are as follows. For $Q_{9}^{+},$ with the values $\beta = 1,\omega = -2,\lambda = -2,\mu
   = 1$ we have  $\left\{-\frac{6}{11},4 \sqrt{\frac{3}{11}},-6
   \sqrt{\frac{3}{11}}\right\}$  which mean saddle behaviour. With the values   $\beta = 1,\omega = 2,\lambda = -2,\mu
   = 1$ we have $\left\{-\frac{6}{5},-\frac{1}{\sqrt{5}},-\frac{1}{\sqrt{5}}\right\}$ this means that it has saddle behaviour (recall, $\rho'$ must be positive for the stability along the radial direction);  For $Q_{10}^{+},$ with the values $\beta = 1,\omega = -2,\lambda = -2,\mu
   = 1$ we have  $\left\{-\frac{6}{11},-4 \sqrt{\frac{3}{11}},6
   \sqrt{\frac{3}{11}}\right\}$  which mean saddle behaviour. With the values   $\beta = 1,\omega = 2,\lambda = -2,\mu
   = 1$ we have $\left\{-\frac{6}{5},\frac{1}{\sqrt{5}},\frac{1}{\sqrt{5}}\right\}$ showing source behaviour.
   
In addition, parameter $v$ should always be real and have a positive value.
Thus, for points $Q_{i}^{+}$ we calculate $v\left(
Q_{i}^{+}\right)  =1$ for $i=1,\ldots, 8$, while $v\left(
Q_{9,10}^{+}\right)  =\sqrt{\frac{\beta  \lambda  (\lambda +4)}{-4 \beta  \lambda +\lambda ^2+\mu ^2 (\omega +1)}+1}$, which means that
they exist for $\frac{\beta  \lambda  (\lambda +4)}{-4 \beta  \lambda +\lambda ^2+\mu ^2 (\omega +1)}>-1$.

The effective equation of the state parameter and deceleration parameter in the Poincar\'e variables are
expressed as
\begin{equation}
w_{eff}\left(  \rho,\theta_{1},\theta_{2}\right)  =-1- \frac{2\rho^2}{1-\rho^{2}}  \sin^2\theta_1 \left(\omega\cos^2\theta_{2}+\beta \sin^2\theta_{2}\right) +\frac{8}{\sqrt{3}}\frac{\rho}{\sqrt{1-\rho^{2}}}\sin\theta_{1}\cos\theta_{2},
\end{equation}

\begin{equation}
q\left(  \rho,\theta_{1},\theta_{2}\right)=\frac{1-\rho ^2+3 \beta  \rho ^2 \cos
   ^2\theta_{1}}{-1+\rho
   ^2}+\frac{\rho  \cos \theta_{2}
   \sin \theta_{1} \left(-4
   \sqrt{3-3 \rho ^2}+3 \rho  \omega  \cos
   \theta_{2} \sin \theta_{1}\right)}{-1+\rho ^2}.
\end{equation} 

Consequently, these values are expected to be divergent when evaluated in the points $Q_i^+$ at infinity; in what follows, we show the leading terms of the series centred on $\rho=1$ to understand the behaviour of the parameters at infinity.
\\
Since $Q_1^+$ and $Q_2^+$ exist for $\omega>0$ we see that $w_{eff}(Q_i^+)\simeq \frac{\omega }{(\rho -1) (\omega
   +1)}$ and $ q(Q_i^+)\simeq \frac{3 \omega }{2 (\rho -1) (\omega
   +1)}$ behave as  $w_{eff}(Q_i^+)\rightarrow (-\infty) \omega,$ $q(Q_i^+)\rightarrow (-\infty) \omega$ for $i=1,2$. Therefore, these points describe Big Rip singularities. For points $i=3,4,5,6$ we have  $w_{eff}(Q_i^+)\simeq -\frac{4 \sqrt{\frac{2}{3}} \sqrt{\beta
   }}{\sqrt{1-\rho } \sqrt{\beta -\omega
   }}-1,q(Q_i^+)\simeq -\frac{2 \sqrt{6} \sqrt{\beta
   }}{\sqrt{1-\rho } \sqrt{\beta -\omega
   }}-1.$ Clearly, for $\beta>0$ and $\beta-\omega>0$ we see that the parameters behave as $w_{eff}(Q_i^+)\rightarrow -\infty$ and $q(Q_i^+)\rightarrow -\infty$, also describing Big Rip singularities.

   For $i=7,8$ we have $w_{eff}(Q_i^+)\simeq\frac{\beta  \omega  \left(16 \beta +\mu
   ^2 \omega \right)}{(\rho -1) \left(16 \beta
   ^2+\mu ^2 \omega ^2\right)}-\frac{16 \beta
   }{\sqrt{1-\rho } \sqrt{24 \beta ^2+\frac{3
   \mu ^2 \omega ^2}{2}}},q(Q_i^+)\simeq\frac{3 \beta  \omega 
   \left(16 \beta +\mu ^2 \omega \right)}{(\rho
   -1) \left(32 \beta ^2+2 \mu ^2 \omega
   ^2\right)}+\frac{1}{\sqrt{\rho -1}}.$ For these points, the parameters behave as $w_{eff}=q(Q_i^+)\rightarrow (-\infty )\beta  \omega  \left(16 \beta +\mu ^2
   \omega \right)$ they describe Big Rip singularities.

   For the final two points, that is $i=9,10$ we have $w_{eff}(Q_i^+)\simeq \frac{\beta  \lambda ^2+\mu ^2 \omega
   }{(\rho -1) \left(-4 \beta  \lambda +\lambda
   ^2+\mu ^2 (\omega +1)\right)},$ and $q(Q_i^+)\simeq \frac{3
   \left(\beta  \lambda ^2+\mu ^2 \omega
   \right)}{2 (\rho -1) \left(-4 \beta  \lambda
   +\lambda ^2+\mu ^2 (\omega
   +1)\right)}$ they behave as $w_{eff}(Q_i^+)=q(Q_i^+)\rightarrow(-\infty ) \left(\beta  \lambda ^2+\mu ^2 \omega
   \right)$ they also describe Big Rip singularities.

\begin{figure}
    \centering
    \includegraphics[scale=0.75]{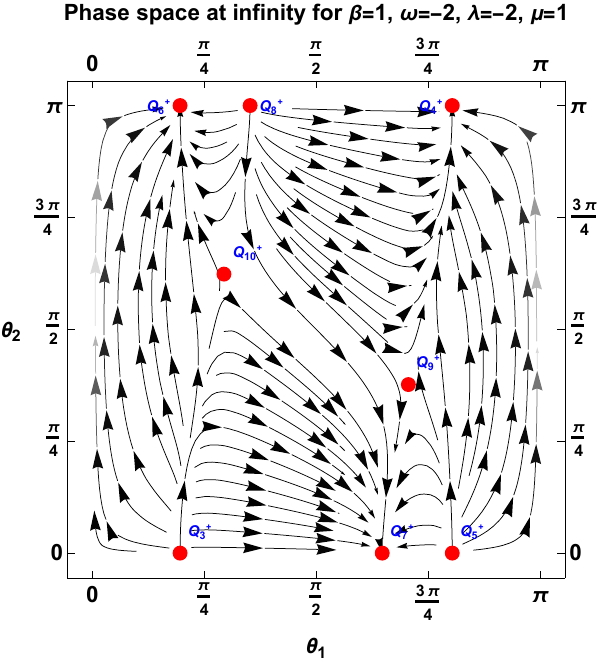}
    \includegraphics[scale=0.35]{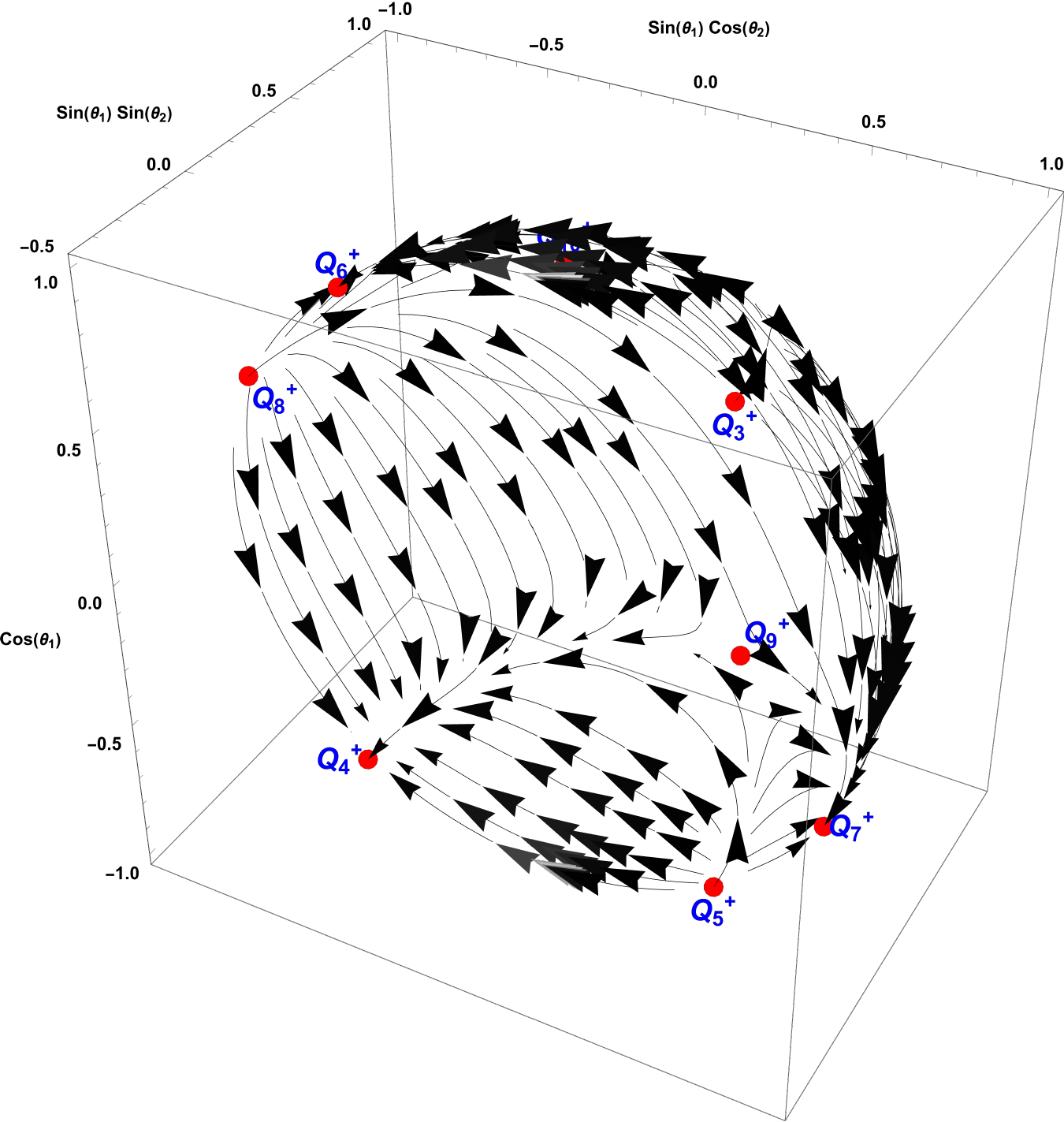}
    \caption{Eight points exist for this choice of parameters. Left: two-dimensional dynamics at infinity ($\rho\rightarrow 1$). Right: The same behaviour is represented in the right hemisphere of the Poincaré sphere.}
    \label{fig-a}
\end{figure}

\begin{figure}
    \centering
    \includegraphics[scale=0.75]{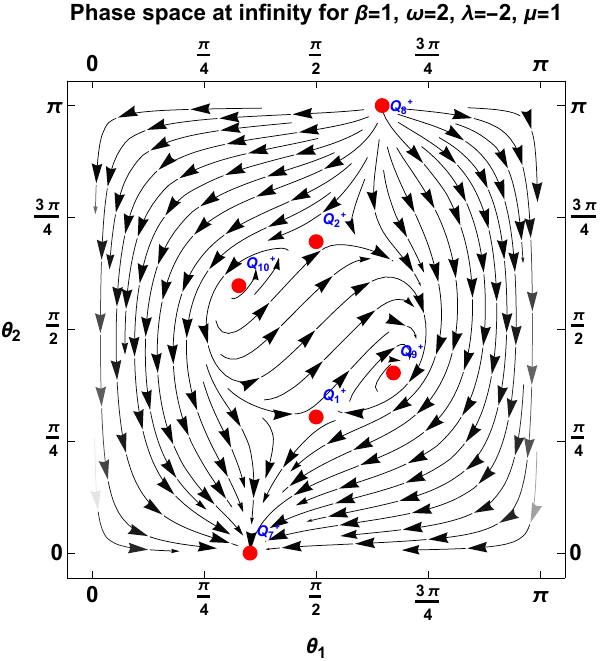}
    \includegraphics[scale=0.35]{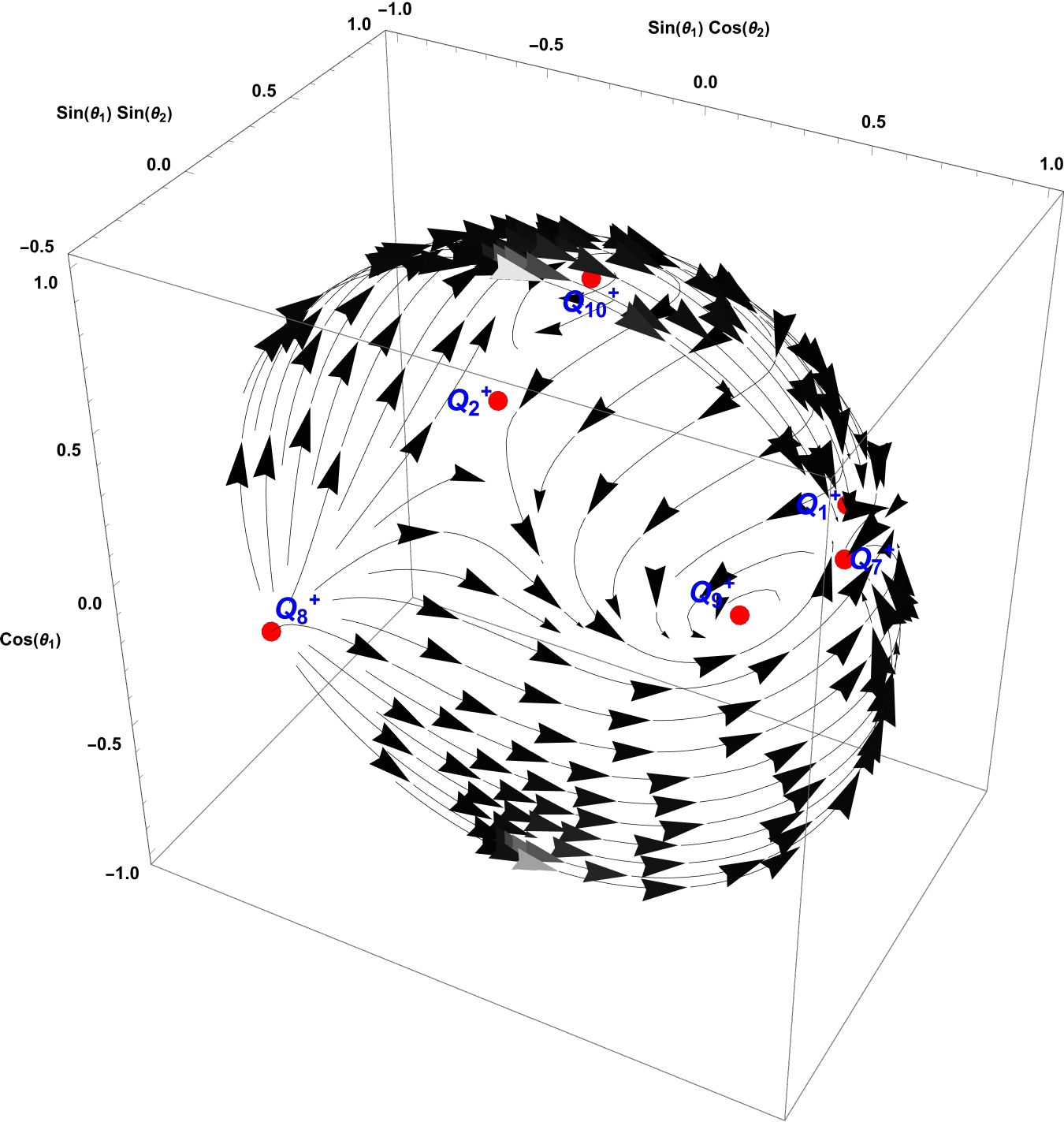}
    \caption{Six points exist for this choice of parameters. Left: two-dimensional dynamics at infinity ($\rho\rightarrow 1$). Right: The same behaviour is represented in the right hemisphere of the Poincaré sphere.}
    \label{fig-b}
\end{figure}

\begin{figure}
    \centering
    \includegraphics[scale=0.75]{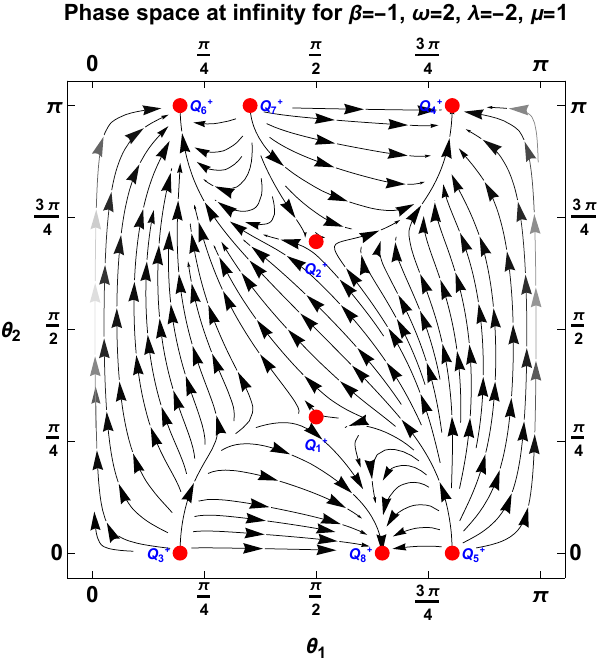}
    \includegraphics[scale=0.35]{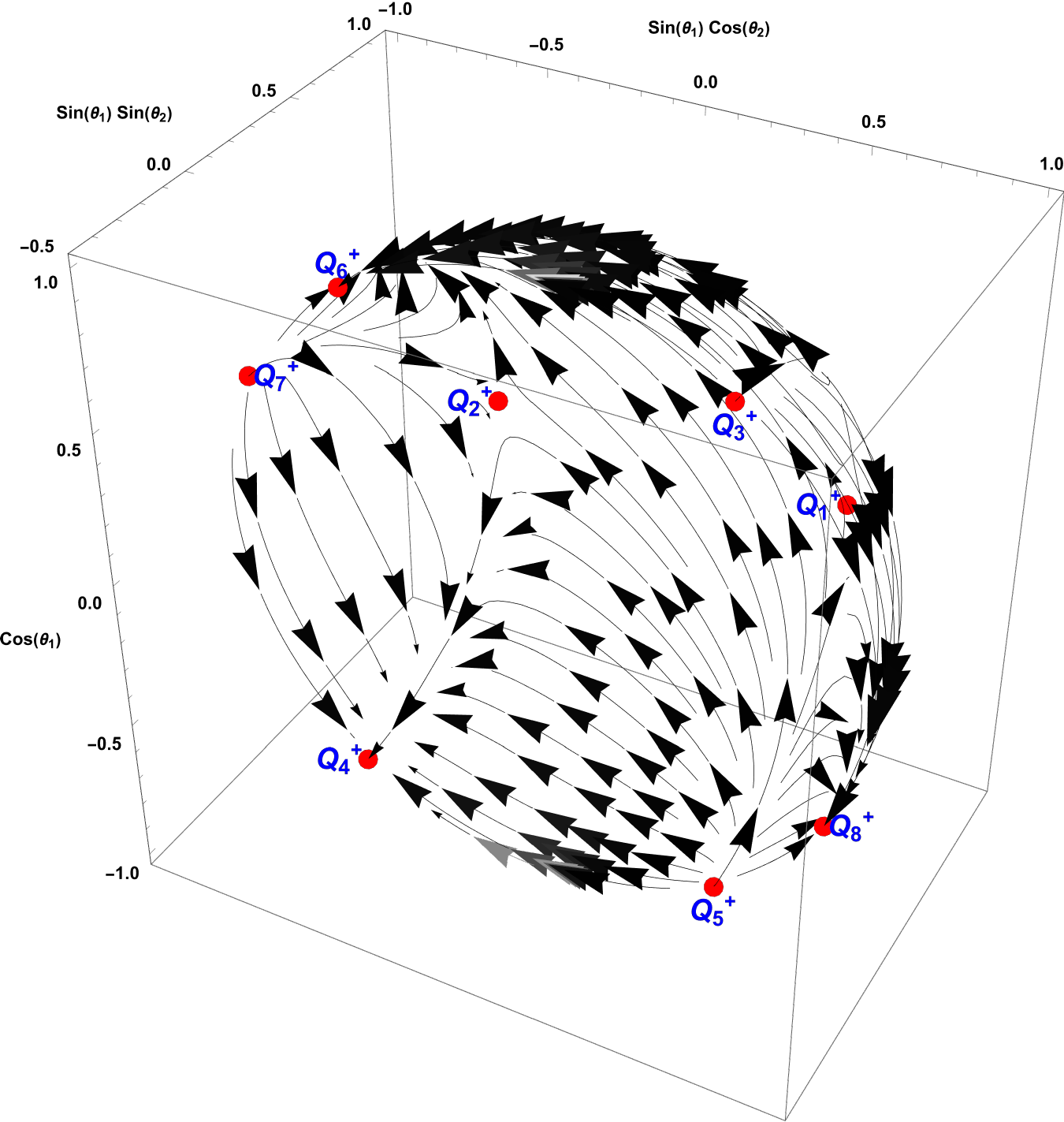}
    \caption{Eight points exist for this choice of parameters. Left: two-dimensional dynamics at infinity ($\rho\rightarrow 1$). Right: The same behaviour is represented in the right hemisphere of the Poincaré sphere.}
    \label{fig-c}
\end{figure}

\begin{figure}
    \centering
    \includegraphics[scale=0.75]{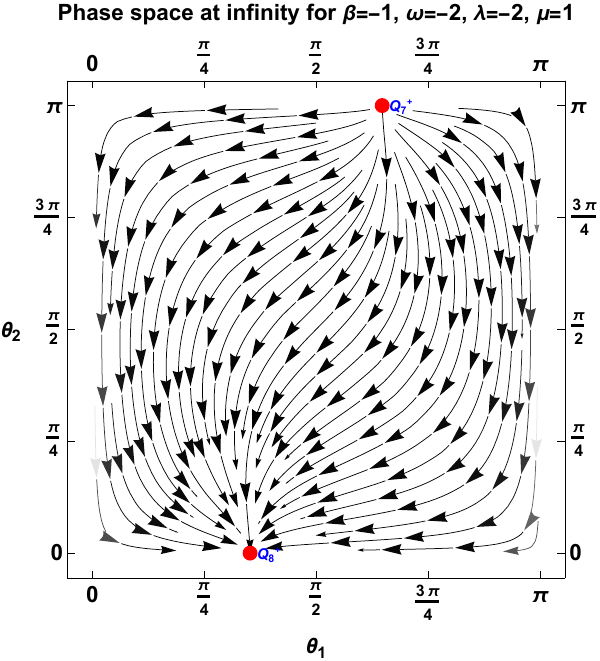}
    \includegraphics[scale=0.35]{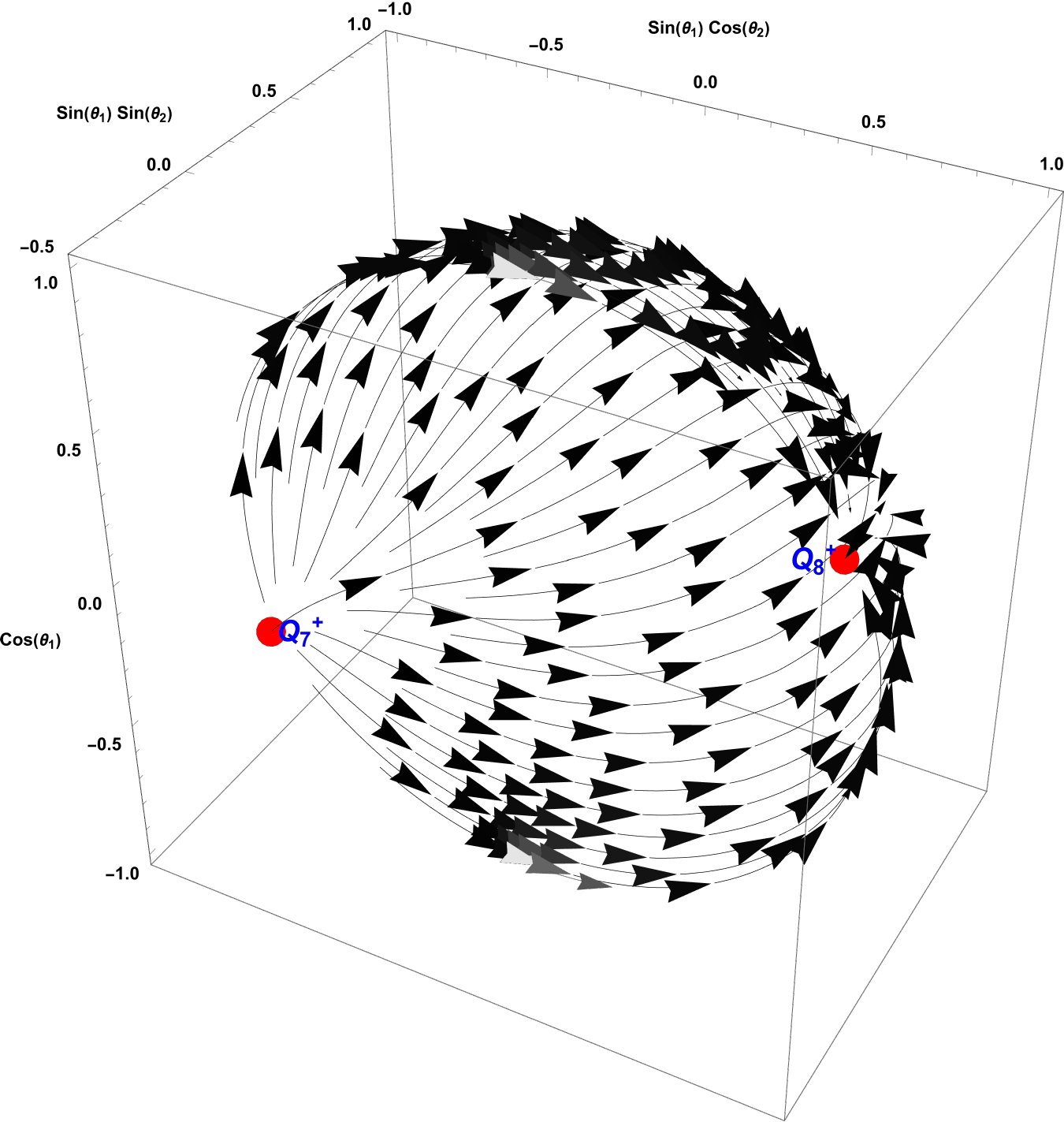}
    \caption{Two points exist for this choice of parameters. Left: two-dimensional dynamics at infinity ($\rho\rightarrow 1$). Right: The same behaviour is represented in the right hemisphere of the Poincaré sphere.}
    \label{fig-d}
\end{figure}

\begin{figure}
    \centering
    \includegraphics[scale=0.75]{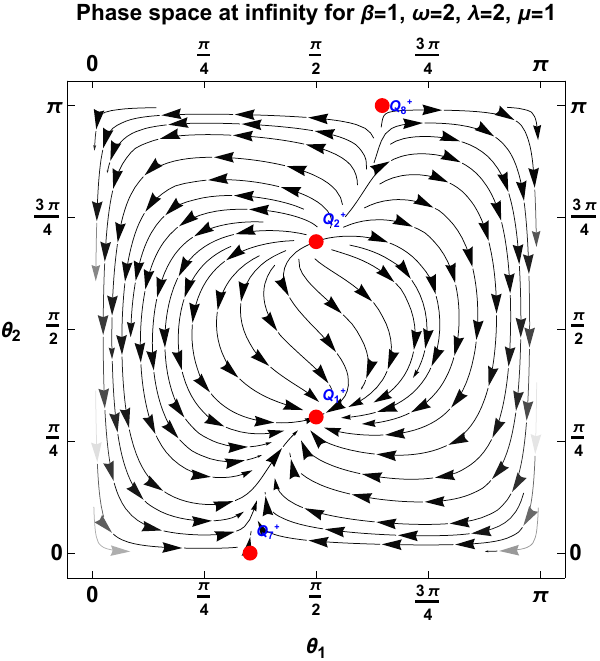}
    \includegraphics[scale=0.35]{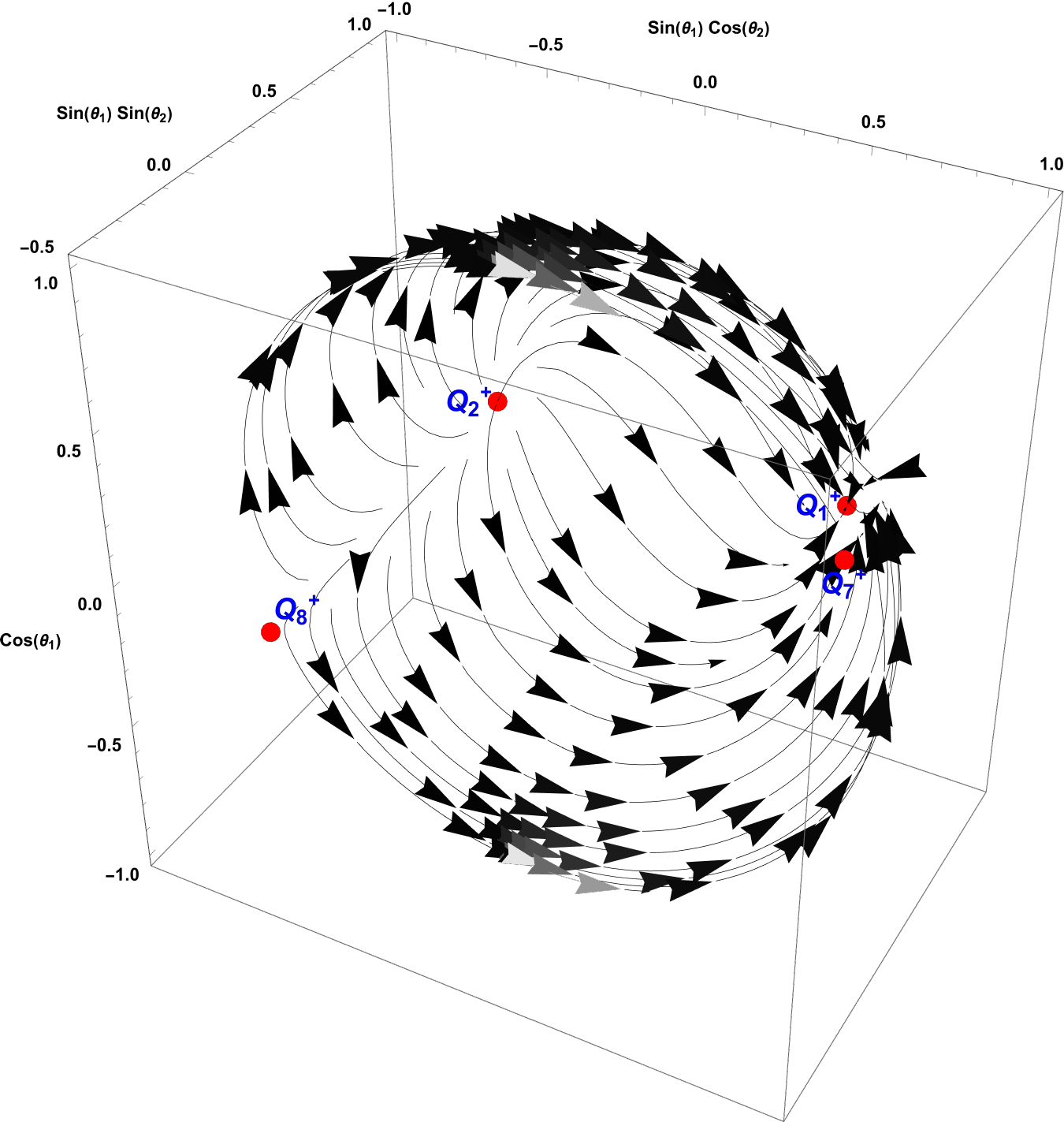}
    \caption{Four points exist for this choice of parameters. Left: two-dimensional dynamics at infinity ($\rho\rightarrow 1$). Right: The same behaviour is represented in the right hemisphere of the Poincaré sphere.}
    \label{fig-e}
\end{figure}

The results of this Section are summarized in Table \ref{tab2}. Moreover, in
Figs. \ref{fig-a}, \ref{fig-b}, \ref{fig-c}, \ref{fig-d}, \ref{fig-e}, we present two-dimensional phase-portraits at the infinity, that is, for $\rho\rightarrow1$, and the same behaviour is shown in the right hemisphere of the Poincaré sphere for  different values of the free parameters.%

\begin{table}[tbp] \centering
\caption{equilibrium points for the multi-torsion cosmological model in the infinity.}%
\label{TII}
\begin{tabular}
[c]{ccccccc}\hline\hline
\textbf{Points} & \textbf{Existence}& \textbf{Attractor?}&$w_{eff}$& $q$ & \textbf{Acceleration}\\\hline
$Q_{1}^{+}$ & see text  & No & $-\infty$&$-\infty$ & Big Rip\\
$Q_{2}^{+}$ & see text  & No&$-\infty$ &$-\infty$ & Big Rip\\
$Q_{3}^{+}$ & see text    & No&$-\infty$ &$-\infty$ &Big Rip\\
$Q_{4}^{+}$ & see text  & No&$-\infty$ &$-\infty$ &Big Rip\\
$Q_{5}^{+}$ & see text   & No&$-\infty$ &$-\infty$ &Big Rip\\
$Q_{6}^{+}$ & see text   & No&$-\infty$ & $-\infty$&Big Rip\\
$Q_{7}^{+}$ & see text   & Yes&$ (-\infty )\beta  \omega  \left(16 \beta +\mu ^2
   \omega \right)$ & $(-\infty )\beta  \omega  \left(16 \beta +\mu ^2
   \omega \right)$ &Big Rip\\
$Q_{8}^{+}$ & see text   & Yes& $ (-\infty )\beta  \omega  \left(16 \beta +\mu ^2
   \omega \right)$ & $(-\infty )\beta  \omega  \left(16 \beta +\mu ^2
   \omega \right)$  &Big Rip\\
$Q_{9}^{+}$ & see text   & No& $(-\infty ) \left(\beta  \lambda ^2+\mu ^2 \omega
   \right)$ &  $(-\infty ) \left(\beta  \lambda ^2+\mu ^2 \omega
   \right)$&Big Rip\\
$Q_{10}^{+}$ & see text   & No& $(-\infty ) \left(\beta  \lambda ^2+\mu ^2 \omega
   \right)$ &  $(-\infty ) \left(\beta  \lambda ^2+\mu ^2 \omega
   \right)$ &Big Rip\\\hline\hline
\end{tabular}
\label{tab2}%
\end{table}%

\section{Conclusions}

\label{conc}

In this study, we explore the dynamic evolution of physical parameters within
a multiscalar gravitational model, specifically in the context of
teleparallelism. Our investigation focuses on the interaction between two
scalar fields that are nonminimally coupled to gravity. By considering a
spatially flat FLRW background and exponential potentials for the scalar
fields, we express the field equations as a three-dimensional first-order
dynamical system.

By analyzing this system, we determine the phase space to reconstruct the
cosmological evolution and understand the history of the physical parameters.
Our analysis involves identifying the equilibrium points and examining the
properties and stability of the corresponding asymptotic solutions.

This particular gravitational model is an extension of the Quintom theory within teleparallelism. The equation of the state parameter can cross the phantom-divide line and the future attractor to provide a universe with a deceleration parameter with a value less than minus one. There are equilibrium points corresponding to asymptotic solutions that describe Big Rip singularities, but these points always act as sources or saddle points. The Big Rip singularities are in the past and describe the very early universe. However, there are also attractors that can explain the present or past acceleration of the universe. Other equilibrium points can represent different eras in cosmological history. Therefore, this gravitational theory provides a framework to unify the components of the dark sector of the universe. These results extend the previous studies of one scalar field \cite{dd0a,dd1}, while new equilibrium points exist where the two scalar fields contribute. The two scalar fields can describe not only acceleration eras of the universe but also attribute other matter components, such as dark matter or radiation, from specific values of the free parameters. For instance, there exists the case where $P_{2}$ or $P_{3}$ can describe a matter-dominated universe.

On the other hand, introducing a
matter component minimally coupled to the scalar fields will not affect the
general behaviour of these equilibrium points. Additionally, to the previous
points, new points which describe the matter are introduced. In this regard, we are looking at the universe's behaviour with respect to its scale factor $a(t)$ under different conditions. When the dark energy equation of state ($w_{eff}$) is not equal to $-1$, the universe's asymptotic solutions have a power-law scale factor: $a(t) = a_0 t^{\frac{2}{3(1+w_{eff})}}$. However, when $w_{eff} = -1$, the universe is described by the de Sitter solution with $a(t) = a_0 e^{H_0 t}$.

To better understand the cosmological history, we also study the stability properties of the equilibrium points. An equilibrium point that is an attractor indicates that the asymptotic solution is stable, while source and saddle points provide unstable solutions. We analyze the stability of these equilibrium points to extract information related to the stability properties of the exact cosmological solutions derived before. We have investigated the evolution of $w_{eff}$ and $q$ evaluated at a numerical solution of equations \eqref{eq-a}, \eqref{eq-b}, and \eqref{eq-c} with initial conditions near $P_2$. Specifically, we took $x(0)=\frac{\lambda +4}{\sqrt{3} \omega }+0.1$, $y(0)=\frac{\sqrt{(\lambda +4)^2+3 \omega }}{\sqrt{3} \sqrt{\omega }}+0.1$, and $z(0)=0.1$. Our analysis shows that for early time, $w_{eff}>-\frac{1}{3}$ and $q>0$; therefore, there is an early-time deceleration phase. Later, the evolution enters a matter-dominated phase when $w_{eff}=0$ and $q=\frac{1}{2}$. Quickly after that, the behaviour describes late-time acceleration for $w_{eff}< -\frac{1}{3}$ and $q<0.$ There is also a transient de Sitter behaviour that occurs when both $w_{eff}$ and $q$ cross the value of -1. Finally, they cross to the phantom regime $w_{eff}<-1$ and $q<-1$ with a transient damped oscillation in both parameters.

We used the Poincar\'e method to extend our study and complete the phase space analysis. We calculated $w_{eff}$ and $q$ at the points $Q_i^+$ at infinity. For $Q_1^+$ and $Q_2^+$, $w_{eff}(Q_i^+)\simeq \frac{\omega }{(\rho -1) (\omega +1)}$ and $ q(Q_i^+)\simeq \frac{3 \omega }{2 (\rho -1) (\omega +1)}$ behave as $w_{eff}(Q_i^+)\rightarrow (-\infty) \omega,$ $q(Q_i^+)\rightarrow (-\infty) \omega$ for $i=1,2$. Therefore, these points describe Big Rip singularities. For points $i=3,4,5,6$ we have  $w_{eff}(Q_i^+)\simeq -\frac{4 \sqrt{\frac{2}{3}} \sqrt{\beta
   }}{\sqrt{1-\rho } \sqrt{\beta -\omega
   }}-1,q(Q_i^+)\simeq -\frac{2 \sqrt{6} \sqrt{\beta
   }}{\sqrt{1-\rho } \sqrt{\beta -\omega
   }}-1.$ Clearly, for $\beta>0$ and $\beta-\omega>0$ we see that the parameters behave as $w_{eff}(Q_i^+)\rightarrow -\infty$ and $q(Q_i^+)\rightarrow -\infty$, also describing Big Rip singularities.  
   For $i=7,8$ we have $w_{eff}(Q_i^+)\simeq\frac{\beta  \omega  \left(16 \beta +\mu
   ^2 \omega \right)}{(\rho -1) \left(16 \beta
   ^2+\mu ^2 \omega ^2\right)}-\frac{16 \beta
   }{\sqrt{1-\rho } \sqrt{24 \beta ^2+\frac{3
   \mu ^2 \omega ^2}{2}}},q(Q_i^+)\simeq\frac{3 \beta  \omega 
   \left(16 \beta +\mu ^2 \omega \right)}{(\rho
   -1) \left(32 \beta ^2+2 \mu ^2 \omega
   ^2\right)}+\frac{1}{\sqrt{\rho -1}}.$ For these points, the parameters behave as $w_{eff}=q(Q_i^+)\rightarrow (-\infty )\beta  \omega  \left(16 \beta +\mu ^2
   \omega \right)$ they describe Big Rip singularities. 
   For the final two points, that is $i=9,10$ we have $w_{eff}(Q_i^+)\simeq \frac{\beta  \lambda ^2+\mu ^2 \omega
   }{(\rho -1) \left(-4 \beta  \lambda +\lambda
   ^2+\mu ^2 (\omega +1)\right)},$ and $q(Q_i^+)\simeq \frac{3
   \left(\beta  \lambda ^2+\mu ^2 \omega
   \right)}{2 (\rho -1) \left(-4 \beta  \lambda
   +\lambda ^2+\mu ^2 (\omega
   +1)\right)}$ they behave as $w_{eff}(Q_i^+)=q(Q_i^+)\rightarrow(-\infty ) \left(\beta  \lambda ^2+\mu ^2 \omega
   \right)$ they also describe Big Rip singularities. To obtain these results, it is necessary to proceed with Poincaré compactification. In Figs. \ref{fig-a}, \ref{fig-b}, \ref{fig-c}, \ref{fig-d}, \ref{fig-e}, we present two-dimensional phase portraits at infinity, i.e., when $\rho\rightarrow1$. The same pattern is shown in the right hemisphere of the Poincaré sphere for various values of the free parameters.

Although the dynamical analysis in a finite regime is complete, it can be extended to cases where the Hubble parameter ($H$) becomes zero. This is because, in some theories, the Hubble parameter changes sign during cosmic evolution and necessarily crosses zero. In such cases, the dynamical variables blow up. In our analysis, we considered $H\neq 0$, and worked on the branch where $H>0$. However, to consider the case where $H=0$, we need to normalize the dynamical variables with the quantity $\sqrt{1+H^{2}}$. This introduces a new variable and increases the dimension of the dynamical system. As the dimension of the dynamical system is already large, we chose to work with the $H$-normalization approach. Nonetheless, we can use alternative methods as described in \cite{Leon:2023ywb,Tot:2022dpr} to analyze the dynamics of gravitational field equations like the Chiral-Quintom theory in a Friedman-Lemaître-Robertson-Walker cosmology with an additional matter source. These methods allow us to study both contracting and expanding universes and to investigate a class of bouncing cosmologies where problems with $H$-normalization typically arise. These methods can help explore the two periods of inflation related to the Universe's early and late-time acceleration phases and the bouncing behaviour. 

In this study, we have focused solely on the case of exponential potentials.
However, in future research, we intend to expand upon this work by considering
more general forms of potential functions. Additionally, we acknowledge the
importance of investigating the evolution of cosmological perturbations in
this background model, and this aspect will be a subject of interest in our
future studies. 

\begin{acknowledgments}
G. L.  was funded by Vicerrectoría de Investigación y Desarrollo Tecnológico (VRIDT) at Universidad Católica del Norte through Resolución VRIDT No. 026/2023 and Resolución VRIDT No. 027/2023 and the support of Núcleo de Investigación Geometría Diferencial y Aplicaciones, Resolución Vridt No. 096/2022.  He also acknowledges the financial support of Proyecto de Investigación Pro Fondecyt Regular 2023, Resoluci\'{o}n VRIDT No. 076/2023. A. P. thanks the support of  VRIDT through Resoluci\'{o}n VRIDT No. 096/2022, Resoluci\'{o}n VRIDT No. 098/2022.
A. D. Millano was supported by ANID Subdirección de Capital Humano/Doctorado Nacional/año 2020 folio 21200837, Gastos operacionales proyecto de tesis/2022 folio 242220121, and VRIDT-UCN. 
\end{acknowledgments}

\end{document}